\documentclass[twocolumn,secnumarabic,amssymb, nobibnotes, aps, prc]{revtex4-2}
\usepackage{graphicx}

\newcommand {\rootsNN}{\mbox{$\sqrt{s_{\mathrm{NN}}}$}}
\newcommand {\roots}{\mbox{$\sqrt{s}$}}

\newcommand{\pT} {p_{\mathrm{T}}}
\newcommand{\jT} {j_{\mathrm{T}}}

\newcommand{\GeV}{GeV}
\newcommand{\TeV}{TeV}

\begin{document}

\title{Novel observables for exploring QCD collective evolution and quantum entanglement within individual jets}%

\author{Austin Baty}%
\email[A. Baty: ]{abaty@rice.edu}
\author{Parker Gardner}%
\email[P. Gardner: ]{ptg2@rice.edu}
\author{Wei Li}%
\email[W. Li: ]{wl33@rice.edu}
\affiliation{Rice University, 6100 Main St., Houston, Texas, 77005}
\date{\today}%

\begin{abstract}
We postulate that non-perturbative QCD effects occurring during parton fragmentation can result in collective effects of a multi-parton system, reminiscent of those observed in high-energy hadronic or nuclear interactions with large final-state particle multiplicity. Proton-proton collisions at the Large Hadron Collider showed surprising signatures of a strongly interacting, thermalized quark-gluon plasma, which was thought only to form in collisions of large nuclear systems. Another puzzle observed earlier in $e^{+}e^{-}$ collisions is that production yields of various hadron species appear to follow a thermal-like distribution with a common temperature. We propose searches for thermal and collective properties resulting from parton fragmentation processes using high multiplicity jets in high-energy elementary collisions. Several novel observables are studied using the PYTHIA 8 Monte Carlo event generator. Experimental observation of such collectivity will offer a new view of non-perturbative QCD dynamics of multi-parton systems at the smallest scales. Absence of any collective effects may offer new insights into the role of quantum entanglement in the observed thermal behavior of particle production in high energy collisions.
\end{abstract}

\maketitle

\section{Introduction}
\label{sec:Introduction}

Quantum Chromodynamics (QCD) is the fundamental theory that 
describes properties of quarks and gluons (known as 
partons) and the interactions among them. Being an $SU(3)$
non-abelian gauge theory, QCD has the peculiar feature 
that partons interact strongly at long distances, 
but become almost free when close to each other 
(``asymptotic freedom'')~\cite{Gross:1973id,Politzer:1973fx,Brock:1993sz}.  As a consequence, no free partons are ever found in the vacuum. 
Instead, partons are always confined inside hadrons.
Attempts to knock a parton out of a hadron and into the 
vacuum (e.g., via a hard scattering in high-energy proton collisions) lead to the emission of collimated sprays of 
hadrons (or ``jets'') that result from fragmentation and 
hadronization of the scattered parton. Detailed dynamics of the parton fragmentation and 
hadronization process are not yet fully understood 
and cannot be evaluated from first-principles because of 
QCD's non-perturbative nature. Phenomenological models such as the Lund string~\cite{Andersson:1983ia}
and cluster models~\cite{Marchesini:1991ch} have been 
implemented to facilitate the
interpretation of experimental data. In recent years, 
there has been tremendous interest and progress in the study
of jet substructures~\cite{Asquith:2018igt}. In these studies,
perturbative QCD (pQCD) approaches
have been successfully applied by largely avoiding or trimming
away non-perturbative components (i.e., soft particles) of the jet~\cite{Asquith:2018igt}.
Fundamental understandings of {\textit{color confinement}} and the
{\textit{dynamics of hadronization}} are two key outstanding 
issues in QCD and strong interactions.

Experiments studying high-energy 
heavy nucleus collisions have been carried out to overcome the QCD confinement and create (possibly thermalized) 
matter with quark-gluon degrees of freedom over an extended 
space-time dimension.
Lattice QCD theory predicts that a crossover
transition to a new phase of partonic matter, known as the
{\textit{quark-gluon plasma}} (QGP), occurs at a temperature of 
about 157~MeV near zero baryon chemical 
potential~\cite{Karsch:1995sy,Karsch:2001cy,Bazavov:2018mes}.
In a high-energy nucleus-nucleus ($AA$) collision (e.g., 
Au or Pb ion), the large volume and density of initial partons 
can lead to many rescatterings, which may rapidly drive the
system toward a thermalized QGP state.
Over the past decades, experiments at CERN's Super 
Proton Synchrotron (SPS)~\cite{CERN-SPS}, BNL's Relativistic
Heavy Ion Collider (RHIC)~\cite{BRAHMS,PHENIX,PHOBOS,STAR}, 
and CERN's Large Hadron Collider (LHC)~\cite{Muller:2012zq} 
have provided compelling evidence for the formation of hot 
and dense QGP matter. Striking long-range collective 
phenomena have been observed and extensively studied
using the azimuthal correlations of particles emitted over 
a wide pseudorapidity range (also known as the ``Ridge'')
at RHIC~\cite{Adams:2005ph,Alver:2008gk,Abelev:2009af,Alver:2009id} and the 
LHC~\cite{Chatrchyan:2011eka,Chatrchyan:2012wg,Aamodt:2010pa,ATLAS:2012at,CMS:2013bza}. These observations indicate that QGP matter is strongly coupled and 
exhibits the hydrodynamic behavior of a nearly perfect liquid~\cite{Ollitrault:1992bk,Teaney:2003kp,Romatschke:2007mq,Heinz:2013th,Gale:2013da}. 

It was thought that elementary collision systems such as 
$e^{+}e^{-}$, proton-proton etc. were too small and dilute 
for any secondary partonic rescattering to occur and drive the system 
toward equilibrium. For this reason, collective flow behavior 
from a QGP medium was not expected in these systems.
Surprisingly, since the start of the LHC, similar long-range
collective azimuthal correlations have been discovered in 
proton-proton (pp) collisions with large final-state 
particle multiplicity~\cite{Khachatryan:2010gv,Aad:2015gqa,Khachatryan:2015lva,Khachatryan:2016txc,Aad:2019aol}, 
which raised the question of whether a tiny QGP droplet with a 
significantly smaller size is created~\cite{Li:2012hc}.
Subsequently, such collective phenomena have been observed in
additional small systems, such as
proton-nucleus ($pA$)~\cite{CMS:2012qk,alice:2012qe,Aad:2012gla,Aaij:2015qcq,ABELEV:2013wsa,Khachatryan:2014jra,Khachatryan:2015waa,Aaboud:2017acw,Aaboud:2017blb,PHENIX:2018lia}
and lighter nucleus-nucleus systems~\cite{Adamczyk:2015xjc,Adare:2015ctn,Aidala:2017ajz,PHENIX:2018lia} 
at RHIC and the LHC.
While it is widely accepted that strong final-state partonic
rescatterings do play a prominent role in the observed 
collectivity of small, high-multiplicity systems, questions 
remain whether the rescatterings are strong enough to drive
the system close to equilibrium or a domain where hydrodynamics
is applicable. 
Meanwhile, alternative scenarios based on gluon saturation 
in the initial state may also contribute, especially at lower 
particle multiplicities (see reviews and latest developments in Refs.~\cite{Dusling:2015gta,Nagle:2018nvi,Schenke:2020mbo}).

It is evident that collective effects of strongly correlated
partonic systems are not only limited to those created in large
$AA$ collisions. Therefore, a series of compelling questions arise: 
\textit{From how small of a system can partonic collectivity emerge and under what conditions? Is partonic collectivity
at such small scales unexpected or a natural consequence of 
QCD in its non-perturbative regime? Can hydrodynamics be
an effective tool in describing non-perturbative QCD dynamics 
of many-body partonic systems (e.g., fragmentation
in the vacuum)?}

In fact, it has been pointed out long ago that total production yields of 
various hadron species in elementary $e^{+}e^{-}$ collisions 
can be well described by a thermal statistical 
model~\cite{Becattini:1995if,Becattini:2008tx,Castorina:2007eb},
similar to that in large AA collisions from a nearly thermalized QGP 
medium~\cite{BraunMunzinger:1994xr}. The origin of this thermal-like
phenomenon in $e^{+}e^{-}$ has not been understood, 
as it was inconceivable that strong final-state partonic 
rescatterings occur there.  There are conjectures that 
thermal-like hadron production is the QCD counterpart of Hawking-Unruh
radiation~\cite{Castorina:2007eb,Hawking:1974sw,Unruh:1976db}.  In recent years, quantum entanglement effects were also proposed to give an intriguing alternative perspective of 
multi-particle production in high-energy collisions
~\cite{Kharzeev:2017qzs,Berges:2017zws,Kaufman_2016,Baker:2017wtt,Tu:2019ouv}. 
The apparent thermalization of final-state hadrons in $e^{+}e^{-}$ 
may be related to dynamics of an expanding quantum string stretched 
between the quark-antiquark pair and its subsequent quenching~\cite{Berges:2017zws}.
No secondary partonic scatterings are involved in this explanation. In these models 
entanglement entropy is calculated with an effective thermal 
temperature and can be related to the temperature extracted
by fitting identified hadron multiplicities to thermal statistical 
models.

Recent experimental searches for long-range ridge correlations 
in $e^{+}e^{-}$~\cite{Badea:2019vey} or $e^{-}p$
collisions~\cite{ZEUS:2019jya} have yielded null results so far, seeming to support the absence of strong final-state rescatterings.  However, in these studies the event multiplicity reach is rather limited 
(up to $\sim$30 tracks in each event), so the presence of rescatterings in small systems achieving higher final-state particle densities can not be ruled out by these data.  As will be discussed further, a single high-multiplicity jet is an example of a small system which may be able to extend the search for final-state rescattering effects to much higher values of local particle density.

Motivated by earlier experimental and theoretical work, our purpose in this paper is to discuss the possibility of understanding the fundamental questions and puzzles 
outlined above from a different view.
\textit{In particular, we postulate that a strongly-interacting, 
QGP-like state\footnote{``QGP-like'' refers to the state where qualitative 
signatures of partonic collectivity are present but the 
system does not necessarily reach the hydrodynamic limit.} 
can indeed originate from a fragmenting quark or gluon as it propagates through the QCD vacuum.}
As a natural consequence of the intrinsic strong QCD coupling strength, the 
strong color fields of the primordial parton in the vacuum can 
give rise to a large number of secondary partons.   
These partons subsequently interact and develop collective expansion which
is transverse to the original parton's direction of motion and extends over a finite space-time volume.
We lay out a proposal to examine a series of key signatures (e.g., 
long-range azimuthal correlations) of such potential QGP-like 
states using energetic jets copiously produced in $pp$ collisions at the 
present CERN-LHC, and also at potential future $pp$, $e^{-}p$ 
and $e^{+}e^{-}$ colliders. Similar studies are also applicable in nuclear 
collisions to explore the ``expansion'' of a parton in a colored medium, instead of the vacuum. Observation of QGP-like signatures 
for a fragmenting parton will provide new insights to the ``thermal'' behavior 
seen from $e^{+}e^{-}$ to $AA$ collisions, and potentially allow a unified view 
of non-perturbative many-body QCD processes (e.g., hydrodynamics models).
Conversely, the 
absence of those collective signatures may highlight the role of quantum 
entanglement effects in parton fragmentation and hadronization.
The direction of research explored by this study shares strong
synergy with both the jet substructure and relativistic heavy ion communities.

The paper is organized in the following way: Section~\ref{sec:parton}
outlines the underlying idea of the possible formation of ``QGP'' from
a parton propagating in the vacuum. Section~\ref{sec:signature}
discusses specific key signatures and how to search for them
experimentally using Monte Carlo (MC) generators for demonstration.
Section~\ref{sec:discussion} is devoted to more physics discussions and extension of proposed studies to other future directions. The paper 
ends with a summary in Section~\ref{sec:summary}.

\section{SINGLE-PARTON ``QGP'' IN THE VACUUM} \label{sec:parton}

\begin{figure}[t!]
\centering
\includegraphics[width=\linewidth]{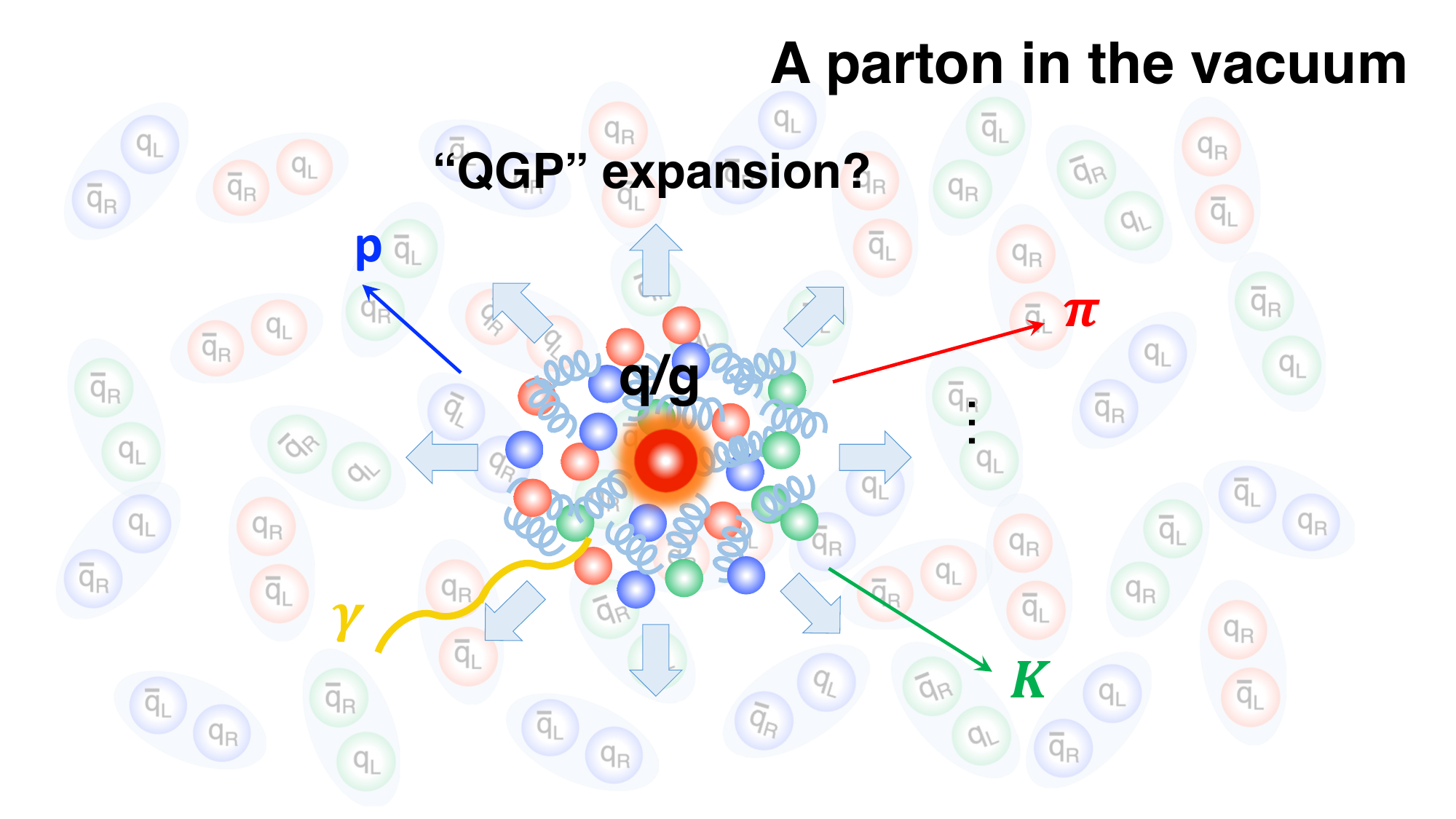}
\includegraphics[width=\linewidth]{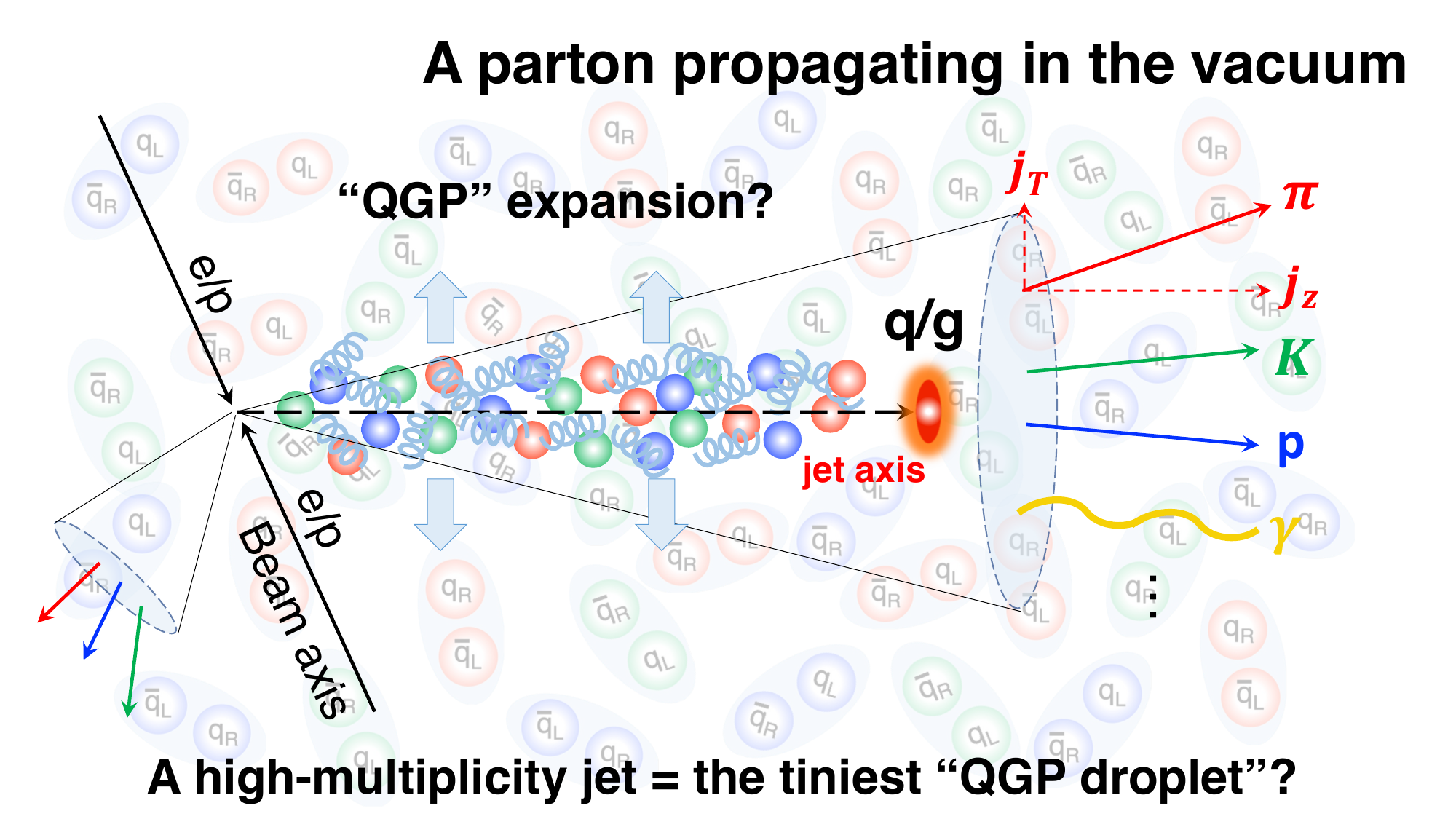}
\caption{Cartoons of a single parton evolving at rest in the vacuum (left) and fast-moving through the vacuum (right).}
\label{fig1}
\end{figure}

In the QCD vacuum state, the chiral symmetry is spontaneously
broken because of the strong coupling nature at low energies.
The QCD vacuum is not empty but filled with non-vanishing condensates 
of quark-antiquark pairs 
($\left< \bar{q}_{L}q_R \right>$+$\left<\bar{q}_{R}q_L \right>$) and gluons 
($\left< G_{\mu\nu}G^{\mu\nu}\right>$), or chiral condensates.  As a thought-experiment, 
consider an extreme (although unrealistic) situation, where
a single parton is placed at rest in the vacuum, as illustrated in Fig.~\ref{fig1} 
(top). The potential energy of associated color fields is infinite. Consequently, more quark-antiquark pairs and gluons will be immediately excited out of the surrounding condensate sea. Thus, in the vicinity of the original isolated parton, a local, dense partonic system will be formed.  Indeed, it may be possible that so many partons will be excited from the vacuum that many of them will have a significant overlap with each other in space and time, which may allow for strong rescatterings between partons. If such rescatterings occur, this may lead to a collective expansion of the partons in the system as they `explode' away from the initial source of high energy densities.  Obviously, the kind of initial conditions depicted in Fig.~\ref{fig1} (top) are not possible to set up experimentally, and a robust theoretical treatment of such a system is hindered by the initial assumption resulting in infinite potential energy in the color fields.  Despite this, we remark that the qualitative behavior conjectured here would be strongly reminiscent of some current models of collective the evolution of a QGP liquid created in a nuclear collision.

Now let us consider more realistic scenarios, where a parton (or partons) 
is knocked out of a proton into the vacuum in hard scattering 
processes of $pp$ collisions.  In the conventional understanding of such a process, the struck parton is highly virtual and can be treated as an essentially free parton because of QCD factorization. However, it immediately begins to shed this virtuality via fragmentation into additional partons in processes that can be calculated by perturbative QCD (pQCD). For example, it is shown by Mueller in Ref.~\cite{Mueller:1982cq} that pQCD calculations can describe \textit{energy dependence} of hadron multiplicity in $e^{+}e^{-}$ collisions producing jets (although it cannot calculate the absolute scale of hadron multiplicity, which is non-perturbative).
Once the virtuality scale approaches $\sim1$~GeV, non-perturbative effects become dominant and the system eventually hadronizes into a shower of final-state particles via inherently non-perturbative processes. Phenomenological models of hadronization, such as the Lund string model~\cite{Andersson:1983ia}, are often employed to describe the remaining details of non-perturbative dynamics. There, excited QCD string systems will create $q\bar{q}$ from the strong field through Schwinger mechanism and form hadrons in the final state~\cite{Andersson:1983ia}. While the string picture is successful in characterizing many
aspects of the parton fragmentation and hadronization process, it still has limitations such as its dependence on many tuning parameters, and its inability to describe the ``thermal behavior'' observed in $e^{+}e^{-}$ collisions.

We can construct a thought-experiment similar the one previously described for an isolated parton, but let the parton be Lorentz-boosted to a fast-moving frame, as illustrated in Fig.~\ref{fig1} (lower).  The scenario is evocative of a similar situation as the beginning stages of a jet's evolution.  In this alternative picture, as the initial parton propagates through the vacuum, its strong color fields will similarly excite (anti-)partons along its path. These excitations develop at the cost of the original parton's energy.  In the case of very high-multiplicity final states, it is plausible to imagine that the created parton (or string) densities are strong enough to result in overlaps and rescattering effects, which could manifest themselves as a collective expansion. In this scenario, the expansion would be most pronounced in the direction transverse to the initial direction of the propagating parton, creating a pattern of partons which is evocative of a ``jet''.   We should stress that a key motivation for proposing this alternative
picture of a parton fragmenting is to try to capture features of non-perturbative processes which are not described by the conventional picture (unless rescattering effects are implemented).  These processes may have connection to the apparent thermal behavior of hadrons in $e^{+}e^{-}$ collisions.

In the conventional picture, the initiating parton starts with a large virtuality which monotonically decreases. However, an isolated parton propagating would start in an on-shell state and gain virtuality through the excitation of additional partons. Nonetheless, as a conventional parton shower evolves towards a collection of low-virtuality partons around the non-perturbative scale, we postulate that certain aspects of systems having many rescatterings and/or collective expansion may also be present in the non-perturbative dynamics of the jet's evolution, similar to the thought-experiment previously described.  If this is the case, we believe high-multiplicity jets, corresponding to a large number of produced partons --- and presumably a higher probability of strong rescatterings in this non-perturbative picture --- would be the best opportunity to study such behavior.  Likewise, for a more dilute system of partons such as a low-multiplicity jet, we expect any rescattering effects to be small enough that existing pQCD and phenomenological models will be sufficient to describe the dynamics of the system.

Experimentally, the process of parton fragmentation 
into hadrons has been studied extensively at colliders.  Recent studies of jet substructure~\cite{Asquith:2018igt}, 
have offered new insights into our understanding of 
the parton fragmentation process, but many of the techniques developed 
there are based on the pQCD and tend to trim away soft-radiated particles 
where intriguing non-perturbative QCD phenomena may occur. Thus, our goal in 
this paper is to focus on studying soft particle production with 
respect to the jet axis, with particular emphasis on
searching for signatures of thermalization and 
collective expansion effects (such as radial and 
elliptic flow as will be discussed in detail later).
We postulate that these effects are particularly likely to develop in high-multiplicity 
jets, where there may be the possibility of creating a system that is characterized by many rescatterings.

\section{SEARCH FOR ``QGP-LIKE'' SIGNATURES WITHIN INDIVIDUAL JETS}\label{sec:signature}

\begin{figure}[t!]
\centering
\includegraphics[width=0.9\linewidth]{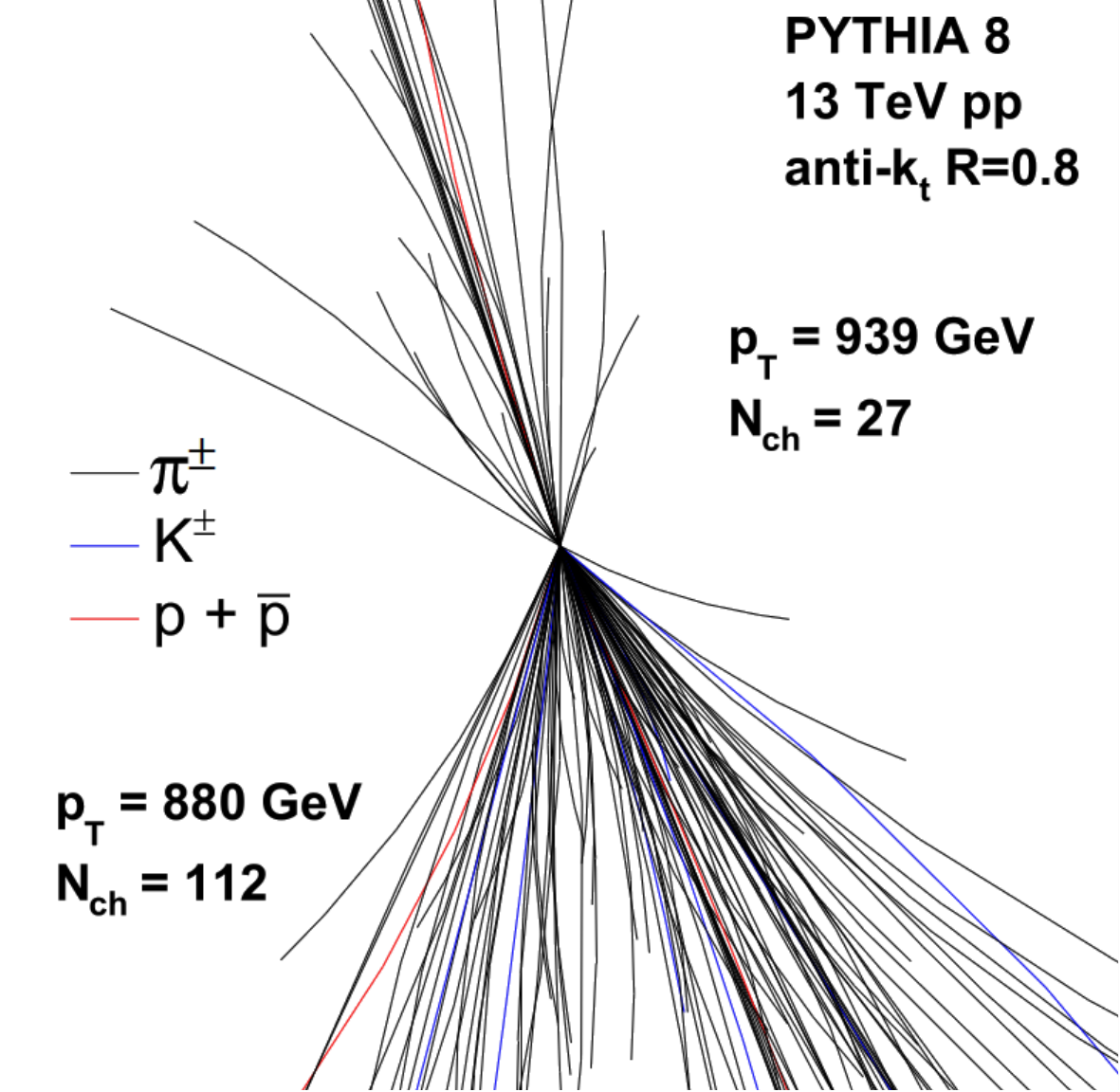}
\caption{A simulated event display for a PYTHIA 8 dijet event in the 
transverse plane with a weak axial magnetic field. One 
of the two jets has a high multiplicity of charged particles 
($N^{j}_{\mathrm{ch}}$), found by the anti-k$_{t}$ algorithm with a 
cone size of R=0.8. Other particles not assigned to the two main 
jets by the jet finding algorithm (e.g., from the underlying event) 
are not shown.}
\label{evtDisplay}
\end{figure}

The analysis strategy discussed in this paper is universally applicable 
to any high-energy collision system including $pp$, $e^{+}e^{-}$ 
and $e^{-}p$, where energetic jets (often dijets) are copiously 
produced. High transverse momentum (relative to the beam axis) jets are first
reconstructed in an event using a particular algorithm (e.g.,
anti-$k_{t}$~\cite{Cacciari:2008gp}) with a choice of jet cone size. 
For an individual jet we define a new coordinate frame such that 
the $z$-axis is aligned with the direction of jet momentum,
named the {\textit{jet frame}}, as illustrated in Fig.~\ref{fig1}.
Momentum vectors of all particles found within the jet cone are 
then re-defined in this new frame, $\vec{p}^{*}=(\jT,\eta^{*},\phi^{*})$.
Here, $\jT$ is the particle transverse momentum with respect to
the jet axis. By selecting very high $\pT$ jets, effects of 
particles from the underlying event that coincidentally fall
inside the jet cone can be significantly suppressed.
We then propose to study a wide range of key QGP signatures,
observed in $AA$ collisions, for particles produced inside high-$\pT$ jets
under this new frame as a function of charged multiplicity inside the
jet cone, denoted as $N_{\mathrm{ch}}^{j}$. We use PYTHIA 8
Monte Carlo (MC) event generator~\cite{Sjostrand:2007gs} 
to demonstrate the proposed analysis strategy.

We first investigate some basic properties of particles produced
within a jet in the new frame. As an illustration of the type of 
events that are being selected in this analysis, a sample 
PYTHIA 8 dijet event 
of $pp$ collisions at $\roots = 13$~TeV is shown in Fig.~\ref{evtDisplay}, 
in the transverse plane of the lab frame. The display 
perspective is along a weak axial magnetic field, 
which causes the charged particles to bend in arcs.  
The two jets are produced and reconstructed with the
anti-k$_{t}$~\cite{Cacciari:2008gp} algorithm 
of cone size R=0.8, each having $\pT$ of roughly 900~GeV.  
The first jet has a fairly average multiplicity of 27, while the second jet  
has over 4 times as many charged particles, and would be classified 
as `high multiplicity'. Different colors of particle trajectories indicate 
different particle species, such as pions, kaons and protons.
Jets reconstructed with smaller cone sizes (e.g., R=0.4) are also investigated
and show qualitatively similar properties of observables studied
in this paper so we focus on presenting results only for jet cone size of 0.8.  The PYTHIA sample used in this study corresponds to an integrated luminosity of approximately 50 $\textrm{fb}^{-1}$ and is filtered to select events having a minimum invariant transverse momentum ($\hat{\pT}$) of 470 GeV.

\begin{figure}[t!]
\centering
\includegraphics[width=\linewidth]{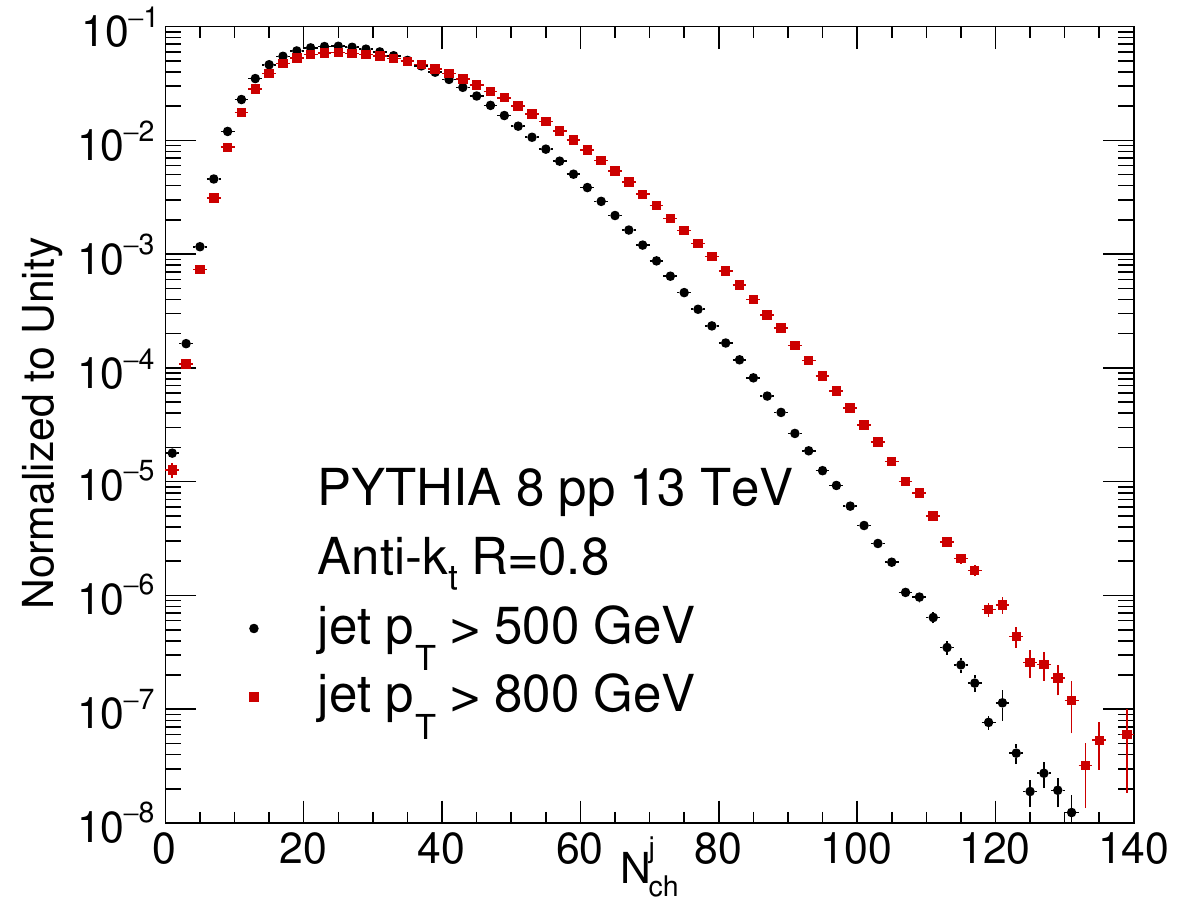}
\caption{Charged multiplicity distributions of jets with $\pT > 500$~GeV
and 800~GeV, respectively. Jets are found by the anti-k$_{t}$
algorithm with a cone size of R=0.8.}
\label{fig2}
\end{figure}

Figure~\ref{fig2} shows multiplicity distributions of 
charged particles within an AK8 (anti-k$_{t}$ reconstruction of cone size R=0.8) jet of $\pT > 500$~GeV 
and $\pT > 800$~GeV in PYTHIA 8. The most likely values of 
charged particle multiplicity are around 25 for both jet $\pT$ selections, 
but the distribution for higher-$\pT$ jets has a slightly longer 
tail at higher multiplicities, reaching up to 140 charged particles.  
This can be understood as the higher jet momentum causes 
more produced particles to fall within the jet cone 
because of the larger Lorentz boost associated 
with the increased parton momentum. However, the difference
between the two selections for a given multiplicity probability 
is only around 10 charged particles at high multiplicities, 
indicating only a loose correlation between jet $\pT$ and 
multiplicity, which is also observed in experimental data 
at sufficiently high jet $\pT$~\cite{Aad:2016oit}. 
Given that high-multiplicity jets are rarely produced and 
that they are only loosely correlated with jet $\pT$, 
suggested by Fig.~\ref{fig2}, standard $\pT$-based experimental 
online triggers are not optimal for studying these systems, 
and a dedicated trigger filtering on high multiplicities 
from a single jet will significantly enhance the potential 
of searching for new phenomena.

\begin{figure}[t!]
\centering
\includegraphics[width=\linewidth]{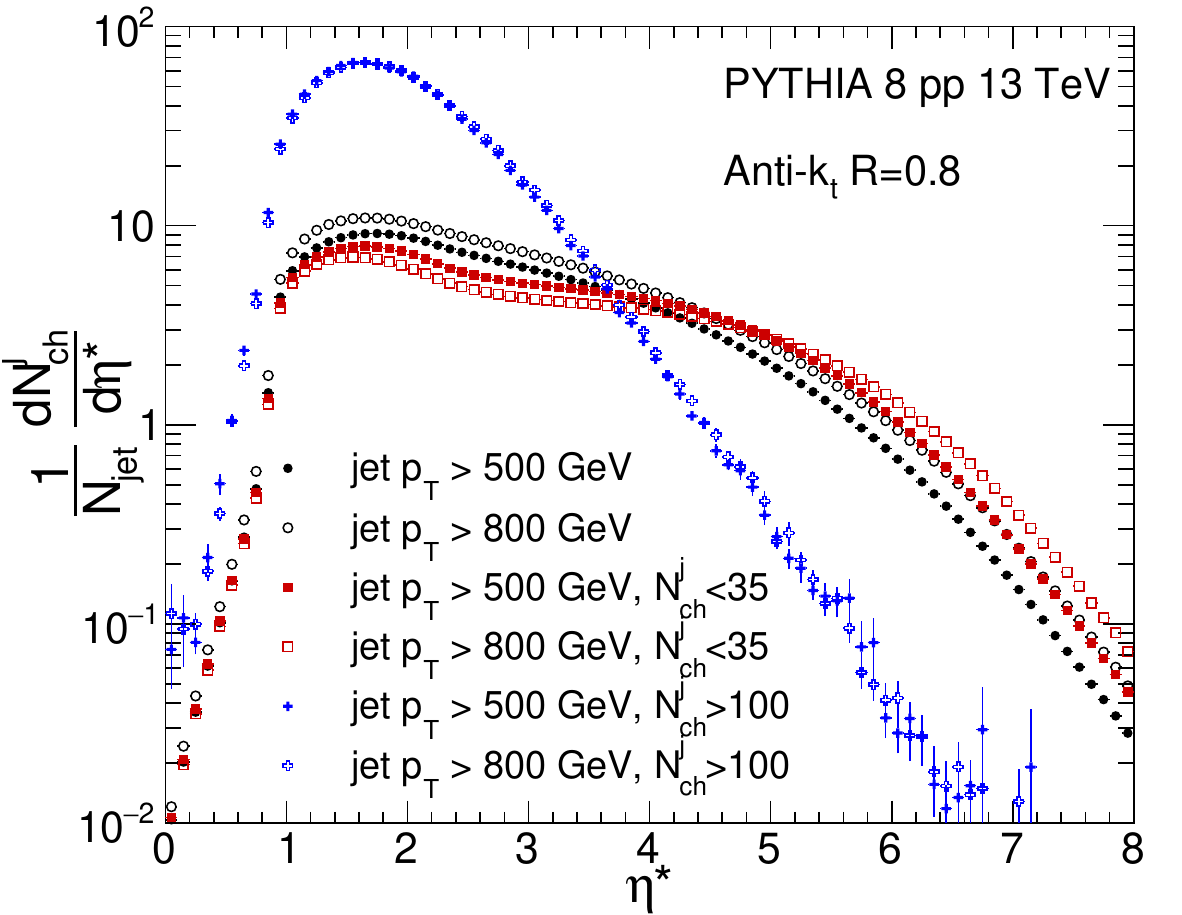}
\caption{The pseudorapidity ($\eta^{*}$) distributions of 
charged particle densities in the single jet frame for 
low ($N^{j}_{\mathrm{ch}}<35$)- and high($N^{j}_{\mathrm{ch}}>100$)-multiplicity 
jets with $\pT >$ 500~GeV and 800~GeV, respectively. 
Jets are found by the anti-k$_{t}$ algorithm with a 
cone size of R=0.8.}
\label{fig8}
\end{figure}

Distributions of charged particle densities in pseudorapidity
of the jet frame, $dN_{ch}/d\eta^{*}$, within 
an AK8 jet are shown in Fig~\ref{fig8}.
In the jet coordinate system, low $\eta^{*}$ corresponds 
to particles that are separated from the main jet axis by 
a large angle, while high $\eta^{*}$ corresponds to particles 
more collimated with the jet direction. The closed 
points show distributions for jet $\pT > 500$~GeV, while 
open points correspond to $\pT > 800$~GeV. An inclusive 
multiplicity selection is shown in black, while 
low ($N^{j}_{\mathrm{ch}}<35$) and high ($N^{j}_{\mathrm{ch}}>100$)
multiplicity selections are shown in red and blue, respectively.  

\begin{figure}[t!]
\centering
\includegraphics[width=\linewidth]{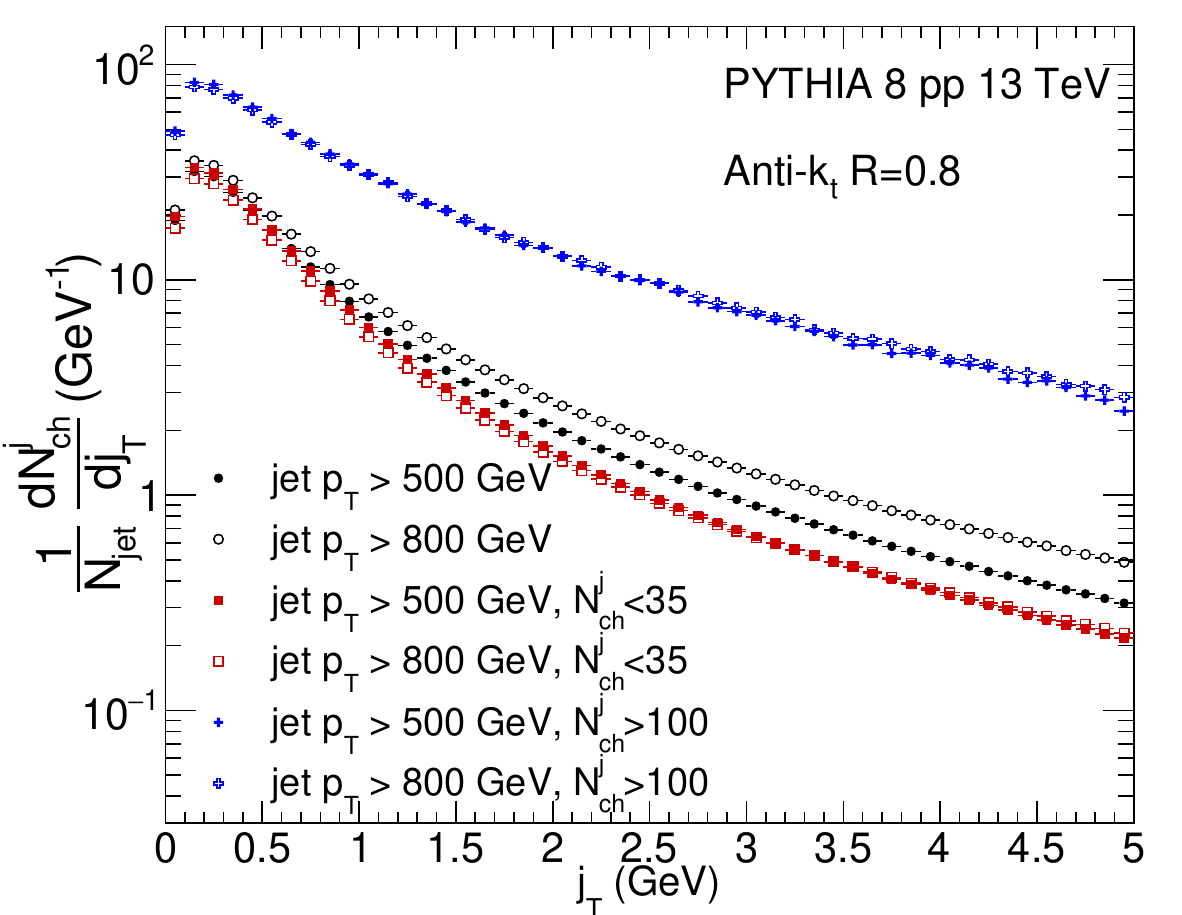}
\caption{The $j_{T}$ distributions of 
charged particles in the single jet frame for 
low ($N^{j}_{\mathrm{ch}}<35$)- and high($N^{j}_{\mathrm{ch}}>100$)-multiplicity 
jets with $\pT >$ 500~GeV and 800~GeV, respectively. 
Jets are found by the anti-k$_{t}$ algorithm with a 
cone size of R=0.8.}
\label{fig9}
\end{figure}

The distribution tends to shift towards lower values of 
$\eta^{*}$, i.e. large emission angles for the high 
multiplicity selection, as compared to the inclusive and 
low multiplicity selections. A similar shift was observed when comparing gluon-initiated jets to quark-initiated jets at the same jet $\pT$.  Thus, a potential explanation for this effect is a correlation between the multiplicity of a jet and the flavor of its initiating parton. 
All three selections have a very sharp rising trend 
around $\eta^{*} = 0.86$ which is related to the angle 
the particle makes with respect to the jet axis being near 
the chosen cone size of 0.8. For the high multiplicity selection, 
the $dN_{\mathrm{ch}}/d\eta^{*}$ reaches values of nearly 70, which 
is comparable to the multiplicity regime where collective effects 
have been observed in high-multiplicity $pp$
collisions~\cite{Khachatryan:2010gv,Aad:2015gqa,Khachatryan:2015lva,Khachatryan:2016txc}.
Therefore, it is feasible to expect that similar multi-parton dynamics 
may be developed inside a jet of sufficiently high multiplicity.
Unlike in $pp$ and $AA$ collisions where there is a wide plateau 
region in dN$_{\mathrm{ch}}$/d$\eta$ over a few units, the $\eta^{*}$ distribution 
of a single jet is much narrower, especially at large multiplicities.
Figure~\ref{fig9} shows $\jT$ distributions of charged particle yields for two different jet $\pT$ selections and multiplicity selections.  All selections exhibit a sharp peak at low $\jT$ values, but the tail of the low-multiplicity selection falls off slightly faster than the inclusive selection.  This is consistent with particles in these jets being emitted as narrower angles relative to the jet axis on average, as was already observed in the $dN_{ch}/d\eta^{*}$ distribution.  The distributions for the high-multiplicity selection are remarkably similar for both jet $\pT$ choices.

In the following subsections, we employ the PYTHIA 8 generator
as a baseline to investigate a series of observables relevant to signatures 
of a QGP and explore potential discoveries in future experiments. 
No effects of rescatterings among produced parton showers or 
strings are expected for the parton fragmentation process in PYTHIA 8 
(or any other MC event generator presently on the market).
The list of observables is not exhaustive but rather representative of
key signatures:
\begin{itemize}
    \item Particle multiplicity and {\textit{strangeness enhancement}} in a dense, thermal partonic medium; 
    \vspace{-0.1cm}
    \item Long-range correlations and {\textit{anisotropy flow}}; 
    \vspace{-0.1cm}
    \item {\textit{Radial flow}} boost to identified particle $\jT$ spectra;
    \vspace{-0.1cm}
    \item {\textit{Quantum Interference}} of identical particles;
\end{itemize}

We shall emphasize that while the ultimate goal is to implement
rescattering (or ``QGP'') effects to the modeling of the parton fragmentation process
to make quantitative predictions of data, we intend to leave that for future work
where dedicated phenomenological efforts and likely some guidance from experimental
data are needed.

\subsection{ Particle multiplicity and strangeness enhancement}
\label{subsec:strange}

We propose to study the total multiplicity of each particle species and their
relative ratios for particles produced from a single jet
in a similar fashion to those in other collision systems. Statistical
models can also be used to describe the particle multiplicity 
data in jets to search for evidence of thermal production from a fragmenting 
parton that may be related to the quantum entanglement effect or a strongly interacting medium. In the analysis presented here, we specifically 
focus on the aspect of strange hadron multiplicities and explore 
possible strangeness enhancement phenomena as a function of the charged 
multiplicity in jets, using the PYTHIA 8 model.

The enhancement of strange hadron production (relative
to non-strange hadrons) in $AA$
collisions has been considered as strong evidence for the
existence of a high-gluon density QGP medium, where 
the gluon splitting channel dominates the strangeness
production~\cite{Koch:1986ud}. In recent years,
it has also been observed that in small $pp$ and $pA$ systems,
strange hadron yields relative to pions smoothly increase as higher
multiplicity events are selected~\cite{ALICE:2017jyt} toward
multiplicity values in $AA$ collisions. 
The PYTHIA 8 model is unable to reproduce the observed 
strangeness enhancement in $pp$ collisions.

\begin{figure}[t!]
\centering
\includegraphics[width=\linewidth]{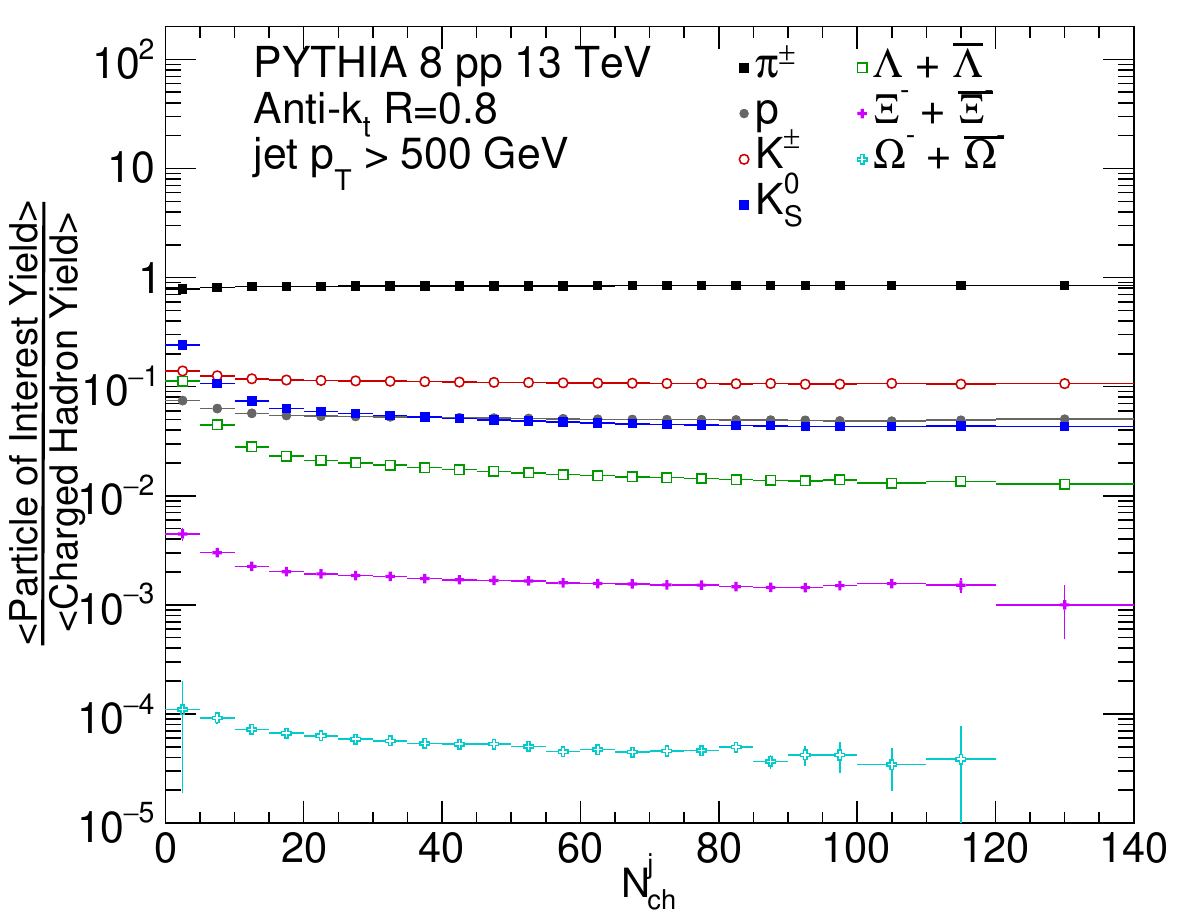}
\caption{Ratios of total yields of various hadrons to 
inclusive charged hadrons from AK8 jets for jet $\pT > 500$~GeV, 
as a function of charged multiplicity in jet ($N^{j}_{\mathrm{ch}}$), 
in PYTHIA 8 $pp$ events at $\roots = 13$~TeV.}
\label{fig6}
\end{figure}

We propose to explore similar strangeness enhancement phenomena 
in high $\pT$ jets, as a function of jet multiplicity.
Using PYTHIA 8 as a reference, the markers in Fig.~\ref{fig6} show 
the ratio of various light and strange hadron yields in 
a high-$\pT$ ($>500$~GeV) AK8 jet to those of charged hadrons, 
as a function of the charged multiplicity of a jet ($N^{j}_{\mathrm{ch}}$).
As expected, no strangeness enhancement is observed in PYTHIA 8.
The ratios of protons to pions is nearly constant as a 
function of $N^{j}_{\mathrm{ch}}$. For strange hadrons such as kaons,
$\Lambda$, $\Xi^-$, and $\Omega^-$, a slight downward trend 
is observed for $N^{j}_{\mathrm{ch}}$ values less than 20, 
but the ratio is nearly independent of multiplicity above this 
threshold. Observation of an increasing strange particle-to-pion yield ratio experimentally in high-multiplicity jets 
would be a compelling indication of additional physics not 
captured by the canonical fragmentation and/or hadronization 
model via string breaking, but possibly involving dynamics 
of dense gluon interactions such as those in high-multiplicity $pp$, 
$pA$ and $AA$ collisions.

\subsection{Long-range correlations and anisotropic flow}
\label{subsec:flow}

Long-range collective phenomena over a wide 
pseudorapidity range have been observed 
in azimuthal correlations of particles from a variety of collision 
systems and experiments. In particular,
the persistence of these collective phenomena in increasingly 
small systems has lead to debates about the origin of 
such behavior and the development of new experiments to push 
the limits of hydrodynamic validity and explore possible effects 
of quantum entanglement.

\begin{figure*}[t!]
\centering
\includegraphics[width=\textwidth]{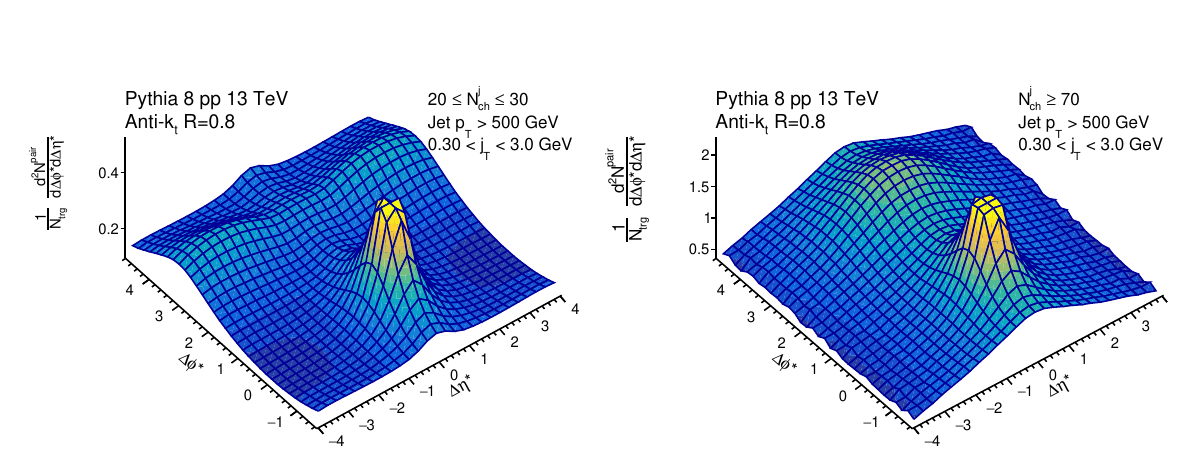}
\caption{The 2-D two-particle angular correlation functions
for particle $0.3<\jT<3$~GeV in low ($20 \leq N^{j}_{\mathrm{ch}} \leq 30$, left)
and high ($N^{j}_{\mathrm{ch}} \geq 70$, right) in-jet
charged multiplicity classes, for AK8 jets with 
jet $\pT > 500$~GeV in PYTHIA 8 $pp$ events at $\roots = 13$~TeV.}
\label{figp1}
\end{figure*}

\begin{figure*}[t!]
\centering
\includegraphics[width=\textwidth]{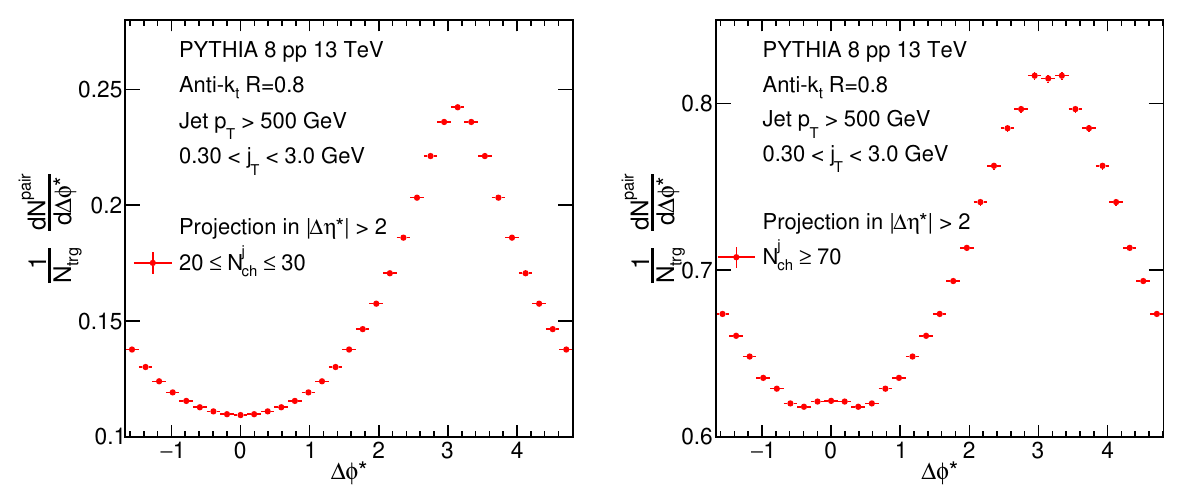}
\caption{The 1-D $\Delta\phi^{\textrm{*}}$ two-particle angular 
correlation functions for particle $0.3<\jT<3$~GeV and 
$|\Delta\eta^{\textrm{*}}|>2$, in low ($20 \leq N^{j}_{\mathrm{ch}} \leq 30$,
left) and high ($N^{j}_{\mathrm{ch}} \geq 70$, right) in-jet
charged multiplicity classes, for anti-$k_{t}$ $R=0.8$ jets with 
jet $\pT > 500$~GeV in PYTHIA 8 $pp$ events at $\roots = 13$~TeV.} 
\label{figp2}
\end{figure*}

We briefly describe the analytical steps of two-particle 
angular correlation analyses in the jet frame and discuss
key features of the result. The procedure is similar to that
employed in Ref.~\cite{CMS:2012qk}, except that the momentum vector of 
all particles are re-defined in the jet frame. 
The 2-D angular correlation function is calculated as follows:
\begin{equation}
    \frac{1}{N_{\textrm{ch}}^{\rm trg}}\frac{\textrm{d}^2N^{\textrm{pair}}}{\textrm{d}\Delta\eta^{*} \textrm{d}\Delta\phi^{*}}=B(0,0)\frac{S(\Delta\eta^{*},\Delta\phi^{*})}{B(\Delta\eta^{*},\Delta\phi^{*})}.
\end{equation}

\noindent where $\Delta\eta^*$ and $\Delta\phi^*$ are relative
pseudorapidity and azimuthal angle in the jet frame, 
for a pair of trigger and associate particles. The trigger
and associate particles can be selected from the same or 
different $\jT$ ranges. For analyses presented below, trigger
and associate particles are chosen from the same $\jT$ range for simplicity. The correlation functions are typically
measured in different $\jT$ and multiplicity ranges.

The $S(\Delta\eta^{*},\Delta\phi^{*})$ and $B(\Delta\eta^{*},\Delta\phi^{*})$ represent the 
signal and background distributions, respectively:
\begin{equation}
    S(\Delta\eta^{*},\Delta\phi^{*})= \frac{1}{N_{\textrm{ch}}^{\rm trg}}\frac{\textrm{d}^2N^{\textrm{sig}}}{\textrm{d}\Delta\eta^{*}\textrm{d}\Delta\phi^{*}},
\end{equation}
\noindent and
\begin{equation}
B(\Delta\eta^{*},\Delta\phi^{*})= \frac{1}{N_{\textrm{ch}}^{\rm trg}}\frac{\textrm{d}^2N^{\textrm{bkg}}}{\textrm{d}\Delta\eta^{*}\textrm{d}\Delta\phi^{*}}.
\end{equation}

\noindent The signal distribution
is calculated with pairs taken from each jet 
($N^{\textrm{sig}}$) and then averaged
over all jets, weighted by the jet multiplicity.
The background distribution serves as a reference
and a correction to the pair acceptance due to 
limited $\eta^{*}$ range. To construct the background 
distribution, we first derive the 2-D single-particle 
$\eta^{*}$-$\phi^{*}$ distribution for daughters of all jets.
Pseudo-particles are then randomly drawn from
the $\eta^{*}$-$\phi^{*}$ distribution 
These pseudo-particles are built from values accumulated 
over multiple distinct jets in multiple distinct events. 
In this way, no correlations should exist 
in the background distribution and all features are detector-related. 
A large number of pseudo-particles $n_{\textrm{pseudo}}$ 
are created such that $N_{\textrm{bkg}} = {n_{\textrm{pseudo}}(n_{\textrm{pseudo}}-1)/2 
\approx 10 \times N_{\textrm{sig}}}$, where 
$N_{\textrm{sig}}$ is the total number of entries in the 
complete signal distribution for the class. The 
$B(0,0)/B(\Delta\eta^{*},\Delta\phi^{*})$ 
term is the appropriate bin-by-bin correction to the 
signal distribution. 

Figure~\ref{figp1} shows the 2-D two-particle angular 
correlation function for low- and high-multiplicity jets 
and particles with $0.3<\jT<3$~GeV in PYTHIA 8 $pp$ collisions at 13 TeV.
The central peak at $(\Delta\eta^{*},\Delta\phi^{*}) = (0,0)$  
is the result of short-range correlations from local parton shower
and hadronization. The far-side ridge 
at $\Delta\eta^{*} \approx \pi$ is mostly related 
back-to-back particle production by conservation 
of momentum. These prominent features 
have been found in lab-frame analyses for both
experimental data and MC simulations. Moreover, 
another feature commonly observed in $AA$ collisions 
is the near-side enhancement at $\Delta\phi^{*} \approx 0$ 
over long-range in $\Delta\eta^{*}$, commonly known as the near-side ``ridge''.  The persistence of this ridge to very small systems, such as $pp$ and $pA$ collisions, naturally motivates a proposal to continue searching for these effects in even smaller systems, like a single jet.
As expected, there is no indication of a near-side ridge for both low- and
high-multiplicity jets in PYTHIA 8. This is also
consistent with $e^{+}e^{-}$~\cite{Badea:2019vey} 
and $e^{-}p$ collisions~\cite{ZEUS:2019jya} at relatively low 
final-state multiplicity. 

The resulting 2-D distribution can be further understood by 
decomposition into a 1-D Fourier series of projections along the 
$\Delta\phi^{*}$ axis:
\begin{equation}
    \frac{1}{N_{\textrm{ch}}^j} \frac{\textrm{d}N^{\textrm{pair}}}{\textrm{d}\Delta\phi^{*}} \propto 1+2\sum_{n=1}^{\infty} V_{\textrm{n}\Delta}(\textrm{j}_{\mathrm{T}}^{\mathrm{A}},\:\textrm{j}_\mathrm{T}^{\mathrm{B}})\cos(\textrm{n}\Delta\phi^{*}), 
\end{equation}

\noindent where $\jT^{\rm A}$ and $\jT^{\rm B}$ represent the
$\jT$ of trigger and associate particles, respectively.
By taking 1-D $\Delta\phi^{\textrm{*}}$ projections over 
$\Delta\eta^{\textrm{*}} > 2$, we exclude the short-range
correlations and focus on understanding structure with large
pseudorapidity separations. The strength of the 
Fourier components in such decomposition can give indications of 
the type of flow and its relative significance in various systems. 
The second Fourier component is typically associated with the 
strength of elliptical flow while the third is associated with the
triangular flow.

In Fig.~\ref{figp2}, 1-D $\Delta\phi^{\textrm{*}}$ 
correlation functions  for $|\Delta\eta^{\textrm{*}}| > 2$
are shown for $20 \leq N^{j}_{\mathrm{ch}} \leq 30$ and 
$N^{j}_{\mathrm{ch}} \geq 70$, respectively, for particles with $0.3<\jT<3$~GeV
from AK8 jets in PYTHIA 8 $pp$ collisions at 13 TeV.
For both multiplicity classes, strong away-side correlations are
observed, consistent with dominant contributions of momentum conservation.
The near-side at $\Delta\phi^{\textrm{*}} \sim 0 $ shows a
minimum, although there seems to an indication of a slight enhancement
for $N^{j}_{\mathrm{ch}} \geq 70$. That enhancement is not significant
and may also be related to the tail of short-range correlations
at very large $\Delta\eta
^{*}$.

\begin{figure}[t!]
\centering
\includegraphics[width=\linewidth]{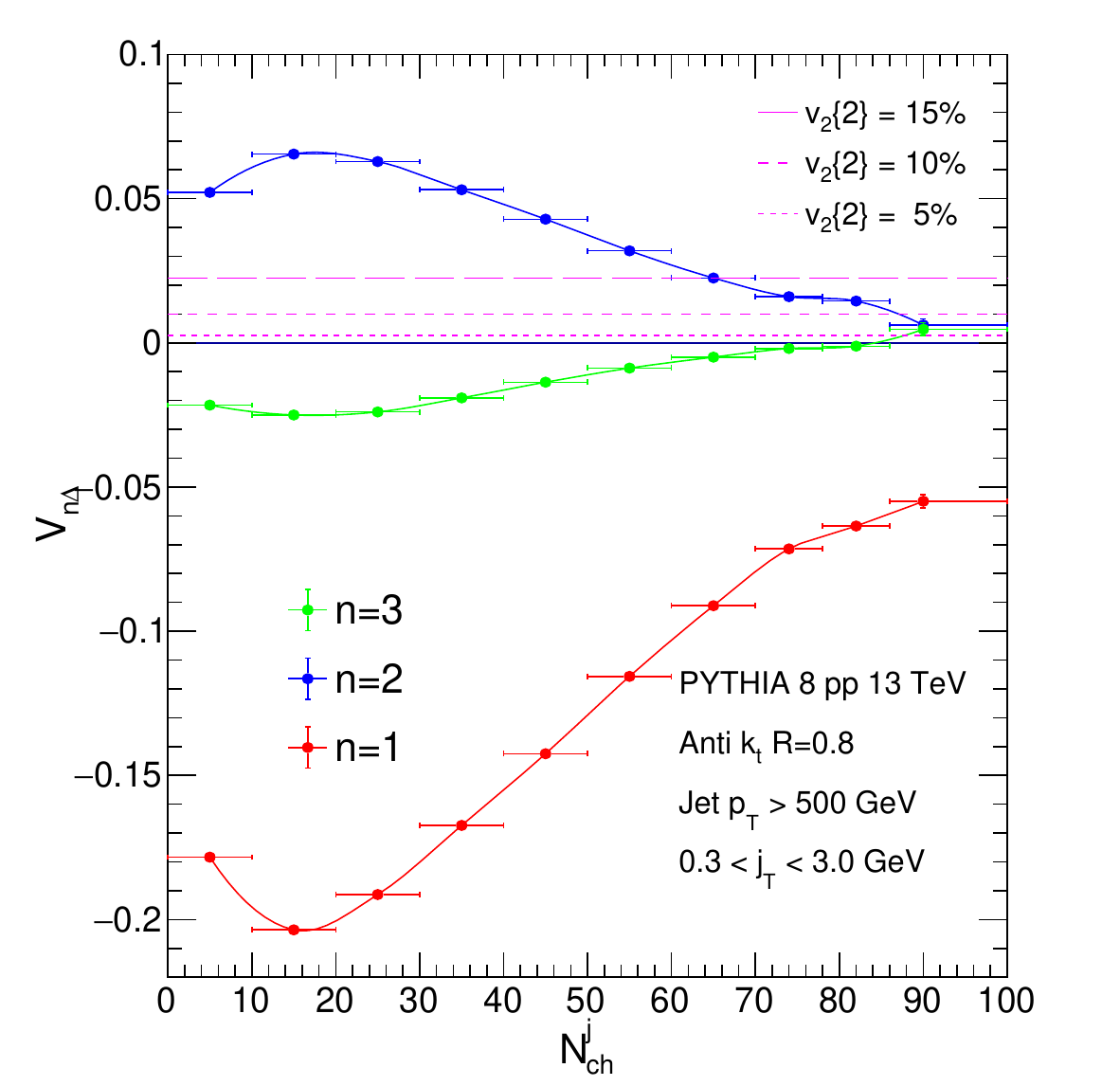}
\caption{The extracted two-particle Fourier
coefficients, $V_{n\Delta}$, as a function of the charged
multiplicity in jet ($N^{j}_{\mathrm{ch}}$), for the first three
harmonic components, from AK8 jets in PYTHIA 8 $pp$ collisions 
at 13 TeV. The dashed lines indicate values of $V_{2\Delta}$ 
equivalent to 5\%, 10\% and 15\% in single particle $v_{2}$.}
\label{figp3}
\end{figure}

The markers and solid lines in Fig.~\ref{figp3} show the extracted two-particle Fourier
coefficients, $V_{n\Delta}$, as a function of the charged
multiplicity in jet ($N^{j}_{\mathrm{ch}}$), for the first three
harmonic components, from AK8 jets in PYTHIA 8 $pp$ collisions 
at 13 TeV. Over the full $N^{j}_{\mathrm{ch}}$ range, the odd-order
harmonics, $V_{1\Delta}$ and $V_{3\Delta}$, are negative, while
the even-odd harmonics, $V_{2\Delta}$ are positive. Magnitudes of
all harmonics decrease as $N^{j}_{\mathrm{ch}}$ increases. All these
features are consistent with expectation of short-range back-to-back
correlations that are not related to collective effects. 
The contribution of short-range few-body correlation to the global
azimuthal anisotropy of the event generally diminishes as 1/$N^{j}_{\mathrm{ch}}$ in the two-particle Fourier coefficients. 
An increase of $V_{2\Delta}$ or a significant positive $V_{3\Delta}$
signal at very high multiplicity could be an indication of the
onset of collective flow effects in the expansion of the parton jet.
Note that the single-particle azimuthal anisotropy Fourier
coefficient, $v_{n}$, is related to the 
two-particle Fourier coefficient 
as $V_{n\Delta}(\jT^{\rm A},\jT^{\rm B})=v_{n}(\jT^{\rm A})v_{n}(\jT^{\rm B})$. Pink dashed lines in Fig.~\ref{figp3}
indicate values of $V_{2\Delta}$ equivalent to 5\%, 10\% and 15\% 
in single particle $v_{2}$. Therefore, if PYTHIA 8 properly models
short-range correlations in the parton fragmentation process, 
an additional 15\% $v_{2}$ enhancement should be clearly identifiable with jets of
$N^{j}_{\mathrm{ch}}>70$, while a much smaller $v_{2}$ enhancement of 5\% would require pushing to much higher multiplicity jets, such as
$N^{j}_{\mathrm{ch}}>$ 90--100. 

\subsection{Identified particle $\jT$ spectra and radial flow}
\label{subsec:spectra}

The hydrodynamic expansion of the QGP will generate a common
velocity field that collectively boosts all produced particles
along the radial expansion direction. This phenomenon is known 
as the ``radial flow'' (see a review in Ref.~\cite{Heinz:2013th}).
As a consequence, final-state particles receive a push to 
higher average transverse momentum, with heavier particles 
gaining more momentum, proportional to the mass. This effect 
can be observed and quantified by measuring the average $\pT$ of 
various particle species in a collision. Besides the average $\pT$,
the average transverse kinetic energy, 
$KE_{T} \equiv m_{T}-m = \sqrt{\pT^2 + m^2} -m$,
is also often used and has the advantage of unifying particle
species of different masses 
(known as the ``$m_{T}$ scaling''~\cite{Gatoff:1992cv}) 
in absence of the radial flow. The $m_{T}$ scaling of hadron 
production in high-energy collisions was proposed 
as early as 1965 by Hagedorn based on a statistical
thermodynamic approach~\cite{Hagedorn:1965st}. 
It has been observed in minimum bias $pp$ collisions,
indicating negligible radial flow effects. In high energy
$AA$ and also high-multiplicity small systems, 
significant breaking of $m_{T}$ scaling is observed 
as the multiplicity or system size
increases~\cite{STAR,PHENIX,Khachatryan:2016yru,Acharya:2020zji,Abelev:2013haa}.

\begin{figure}[t!]
\centering
\includegraphics[width=\linewidth]{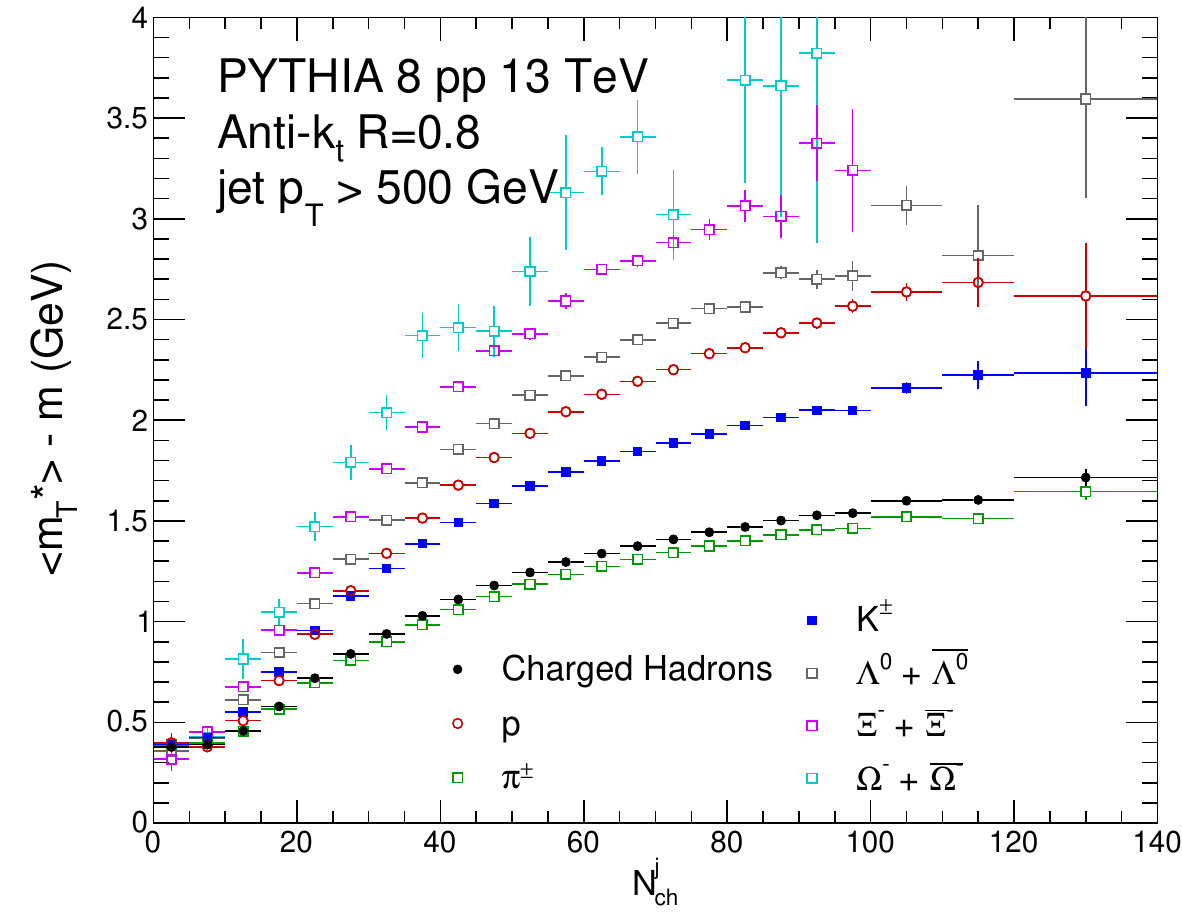}
\caption{The average transverse kinetic energy of various particle species produced
from AK8 jets for jet $\pT > 500$~GeV, as a function of charged 
multiplicity in jet ($N^{j}_{\mathrm{ch}}$), in PYTHIA 8 $pp$ events at $\roots = 13$~TeV.}
\label{fig3}
\end{figure}

We present the average kinetic energy, $\left<m^{*}_{T}\right>-m$, 
calculated in the jet frame in Fig.~\ref{fig3}, 
for inclusive charged hadrons, charged pions, charged 
kaons, protons, $K^{0}_{s}$, $\Lambda$, $\Xi^{-}$ and $\Omega^{-}$ 
produced within AK8 jets, for jet $\pT>500$~GeV as a function of charged
multiplicity in jet ($N^{j}_{\mathrm{ch}}$) in PYTHIA 8 $pp$ events 
at $\roots = 13$~TeV. The $m_{T}$ scaling is indeed present
for low-multiplicity jets ($N^{j}_{\mathrm{ch}}<10$) in PYTHIA 8.
As $N^{j}_{\mathrm{ch}}$ increases, an increasing trend 
of $\left<m^{*}_{T}\right>-m$ is observed for all particle species
but they do not appear to fall on a common trend. Instead, 
heavier particles appear to have greater average kinetic energy
values, and by extension, greater average $j_{T}$ 
values as well. This trend is qualitatively similar to that 
observed in high-multiplicity $pp$, $pA$ and $AA$ collisions. 
The rate of increase of $m_{T}$ with multiplicity seems 
greatest in the range of $N^{j}_{\mathrm{ch}} \sim 20$--30, 
with a flattening trend at higher multiplicities. 
The breakdown of $m_{T}$ scaling as a function of multiplicity 
in PYTHIA 8 is possibly related to the color reconnection effect,
which effectively generates a boost to final-state particles.
Therefore, this observable alone should not be taken as a 
unique signature of the QGP-like state formation. Quantitative
comparison with theoretical calculations, as well as supporting
evidence from other observables, would be necessary to draw a 
conclusion.

\subsection{\normalsize Quantum interference of identical particles}
\label{subsec:bec}

The Bose-Einstein Correlations (BEC), or interferometry exploits 
quantum interference effects of identical particles produced 
with overlapping wave functions in phase space. By studying momentum 
correlations of two identical bosons (fermions), an enhancement 
(depletion) will be observed at small momentum difference between 
two particles. The size of the source of particle emission at 
``freeze out'' (when particles cease to interact) in spacial coordinate 
space can then be inferred from the correlation range in the 
momentum space. The two-particle intensity interferometry
method was first invented by Hanbury, Brown and Twiss (HBT)
to measure the size of astronomical objects~\cite{HanburyBrown:1956bqd}.
It has since been extensively applied to extract the space-time structure of 
QGP in $AA$ collisions~\cite{Zajc:1984vb,Heinz:1999rw,Lisa:2005dd}.

In this study, the two-particle BEC correlation function 
is defined as the ratio,
\begin{equation}
    C_{2}(\vec{q}^{*}) \equiv \frac{S(\vec{q}^{*})}{B(\vec{q}^{*})},
\end{equation}

\noindent where 
$\vec{q}^{*} = \vec{p}^{*}_{1} - \vec{p}^{*}_{2}$ is the
momentum difference between the two particles 
in the jet frame. Similar to two-particle angular correlations, 
the $S(\vec{q}^{*})$ is the measured
particle pair distribution from the same jet containing potential BEC
signals, while the $B(\vec{q}^{*})$ is formed with pairs of
random pseudo-particles as a reference, as well as for correction of
detector effects.
The BEC studies can be performed in one, two or three dimensions
of $\vec{q}^{*}$. For simplicity, we present a 1-D analysis in 
$q^{*}_{\rm inv}=|\vec{q}^{*}|$ using
charged pions from jets in PYTHIA 8 as a function of the 
charged multiplicity in jet ($N^{j}_{\mathrm{ch}}$) and pair transverse momentum, 
$k_{T}^{*} =\frac{1}{2}|\vec{j}_{T,1} + \vec{j}_{T,2}|$, 
to demonstrate the idea. No BEC signals are implemented in the PYTHIA 8
MC generator.

\begin{figure}[t!]
\centering
\includegraphics[width=\linewidth]{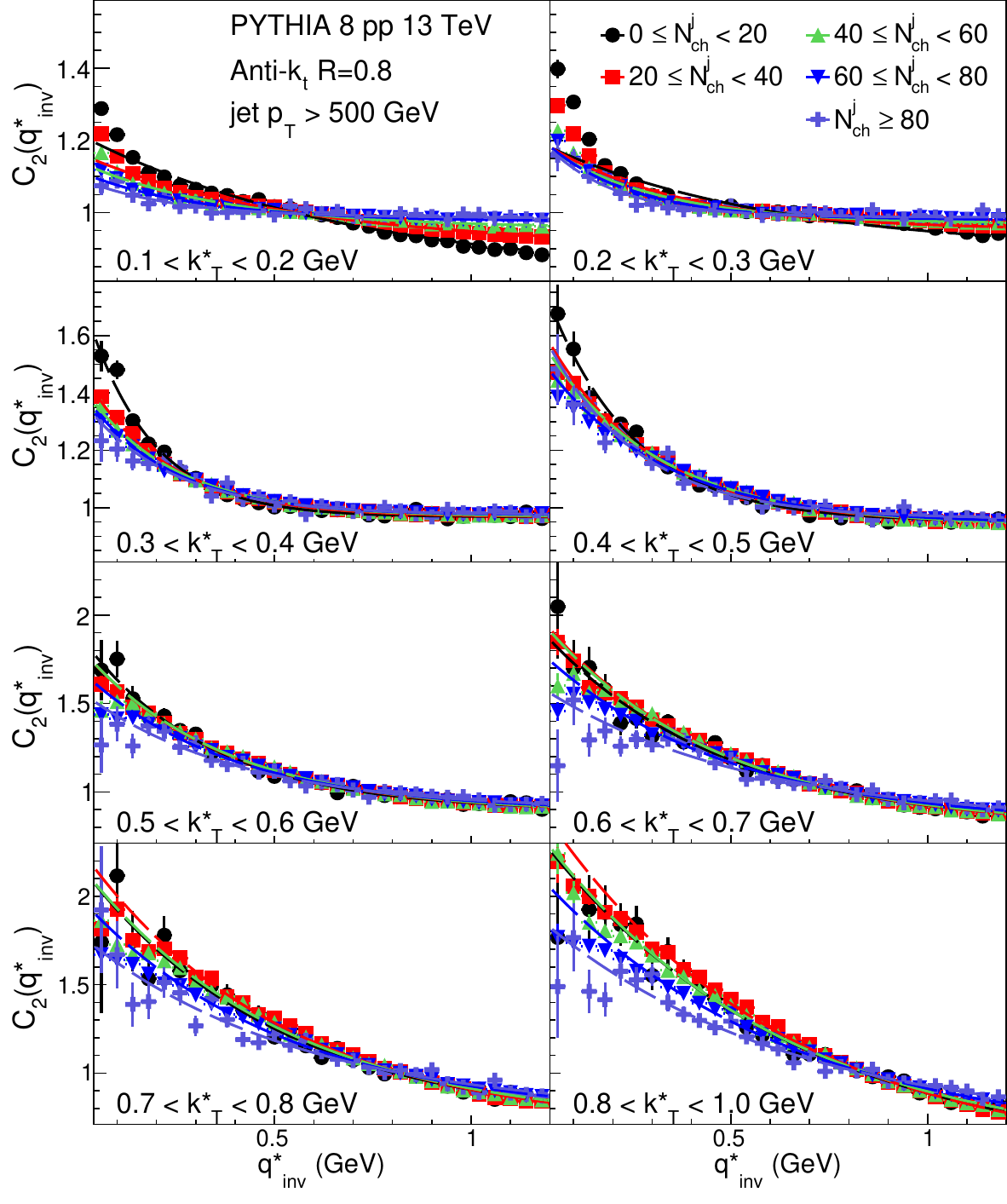}
\caption{The 1-D BEC correlation functions for pairs of same-sign 
charged pions in several ranges of
charged multiplicity in jet, $N^{j}_{\mathrm{ch}}$, for AK8 jets 
with jet $\pT > 500$~GeV in PYTHIA 8 $pp$ 
events at $\roots = 13$~TeV. Each panel represents a range of
pair transverse momentum $k_{T}^{*}$, defined in the single jet frame.
} 
\label{fig:BEC_CF}
\end{figure}

Figure~\ref{fig:BEC_CF} presents 1-D BEC correlation functions 
for pairs of same-sign charged pions in $q^{*}_{\rm inv}$
in different $N^{j}_{\mathrm{ch}}$ ranges of AK8 jets
in PYTHIA 8 $pp$ events at $\roots = 13$~TeV. Each panel of
Fig.~\ref{fig:BEC_CF} shows results for each $k_{T}^{*}$ range. 

All $C_{2}(\vec{q}^{*})$ distributions show a general trend 
of enhanced correlations toward $q^{*}_{\rm inv}\sim0$. 
This feature is qualitatively similar to
those observed in $AA$ collisions~\cite{Lisa:2005dd}, 
where quantum interference effects are believed to play 
the dominant role. The width of $C_{2}(\vec{q}^{*})$ is inversely
proportional to the size of the particle-emitting source. 
As mentioned earlier, there are no BEC correlations
expected in PYTHIA 8. Therefore, these results reflect
the background contributions from the fragmentation.
To distinguish the background contribution from true
BEC signals, it is necessary to investigate the detailed
$N^{j}_{\mathrm{ch}}$ and $k_{T}^{*}$ dependence.

\begin{figure}[t!]
\centering
\includegraphics[width=\linewidth]{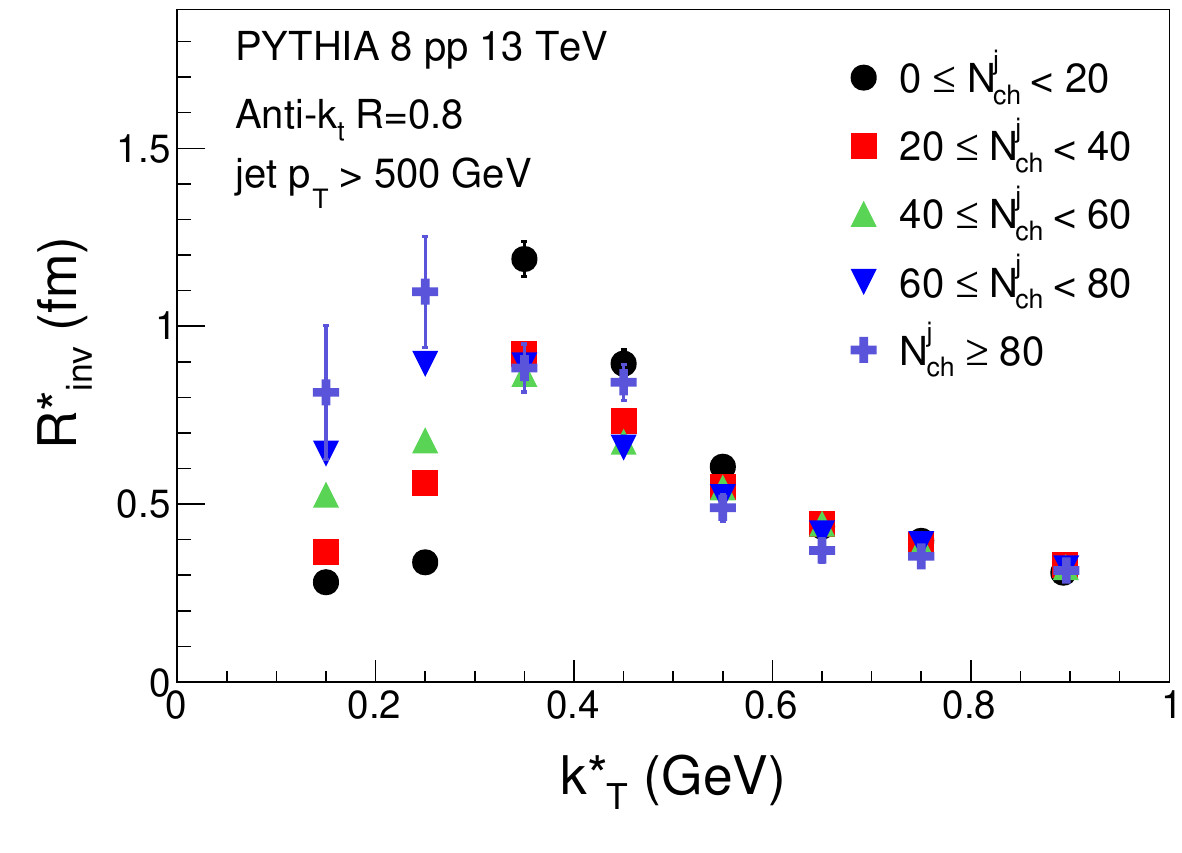}
\caption{The extracted 1-D BEC radii for charged pions as a 
function of pair transverse momentum $k_{T}^{*}$, 
defined in the single jet frame, in several ranges of
charged multiplicity in jet, $N^{j}_{\mathrm{ch}}$, for AK8 jets with 
jet $\pT > 500$~GeV in PYTHIA 8 $pp$ events at $\roots = 13$~TeV.}
\label{fig:BEC_radii}
\end{figure}

The 1-D BEC correlation function is fitted by an 
exponential function:

\begin{equation}
C\left(1+\lambda e^{-R_{\rm inv}^{*}q^{*}_{\rm inv}}\right),
\end{equation}

\noindent where the parameter, $R_{\rm inv}^{*}$, 
characterizes the size of the coherent source (in unit of fm). 
The extracted values of $R_{\rm inv}^{*}$ are shown in
Fig.~\ref{fig:BEC_radii}, as a function of $k_{T}^{*}$, 
for several $N^{j}_{\mathrm{ch}}$ ranges of AK8 jets $\pT > 500$ 
GeV in PYTHIA 8 $pp$ events at $\roots = 13$~TeV. 

In $pp$, $pA$ and $AA$ collisions, the BEC radii parameter
is observed to monotonically increase as the pair momentum 
decreases. This can be understood by the uncertainty 
principle that larger sources tend to coherently emit 
particles at lower momenta. The BEC radii are also found 
to increase with event multiplicity approximately to 
the power of 1/3, which is again consistent with the 
formation of a medium that expands collectively. 
In the BEC analysis of particles in the jet frame using PYTHIA,
shown in Fig.~\ref{fig:BEC_radii}, similar features of extracted
radii in $AA$ collisions are not observed. As seen in 
Fig.~\ref{fig:BEC_radii}, the radii parameter in the jet frame,
$R_{\rm inv}^{*}$, shows a non-monotonic behavior as a function of 
$k_{T}^{*}$, which first increases but then decreases toward 
low $k_{T}^{*}$. While the $R_{\rm inv}^{*}$ does increase 
with $N^{j}_{\mathrm{ch}}$ for $k_{T}^{*}<0.3$~GeV, this trends 
becomes opposite at higher $k_{T}^{*}$. Therefore, a systematic
study of BECs for particles in high-$\pT$ jets as a function of
multiplicity and pair transverse momentum in the jet frame
has the potential to provide key evidence for the formation of
a system with extended space-time structures.

\section{DISCUSSIONS} \label{sec:discussion}

In this paper, we have been focusing on studies of particles 
produced in inclusive single jets with high multiplicity 
in $pp$ collisions at LHC energies. There are many possible 
extensions of proposed analyses in other directions to 
explore new phenomena in high density QCD physics experimentally.
We discuss a few examples below.

{\textit{Thermal Photon Emission:}} In $AA$ collisions, the observation of a large excess of soft 
photons at low $\pT$ ($<1$~GeV) over the primordial hard photon 
production in perturbative QCD
processes~\cite{Adare:2008ab,Adam:2015lda,STAR:2016use} 
is considered as direct evidence for the formation of a 
thermalized QGP medium. The slope of excess photon
$\pT$ spectra provides direct information of the QGP's temperature. 
If such a medium was produced in a high-multiplicity jet system, 
a similar enhancement of photons at small $\jT$ inside the jet cone 
would also be expected. Those photons are typically identified 
as fragmentation photons, emitted from parton showers. 
Even if emitted with small $\jT$, these photons could still have 
a relatively large $\pT$ in the laboratory reference 
frame, and therefore be measured by experiments like CMS and ATLAS, 
which have calorimeters optimized for the high-$\pT$ photon regime.  
However, this type of measurement will have to deal with huge backgrounds originating from hadron (e.g., $\pi^{0}$) decays, as well as underlying event contributions and will be undoubtedly extremely challenging.  At high energy lepton-lepton and 
lepton-hadron collisions, the the underlying event background 
is much cleaner. Therefore, future high-energy $e^{+}e^{-}$ 
colliders and the electron-ion collider planned in the USA may provide 
an ideal environment to search for thermal photon production 
from a single parton.

{\textit{Di-jets and vector boson-jet systems:}} Vector boson-jet 
events, such as Z/$\gamma$-jets, in $pp$ collisions are ideal tools 
to study quark propagation in the vacuum and possible 
collective effects developed around the quark direction of motion. 
All analyses performed with inclusive jets can be done 
with Z/$\gamma$-jets in the same way. A back-to-back di-jet system 
in $pp$ collisions is reminiscent of the final state of $e^{+}e^{-}$ 
collisions, where a color string may be stretched between the 
two fast-moving partons and develop interesting dynamics. 
Correlating particles from two different jets also helps extend 
the rapidity gap of two particles and benefit the
search for long-range correlations. There are some complications
to analyses in the di-jet system though. As the two jets are never 
exactly back-to-back, choosing a common $z$ axis for the new frame 
(e.g., the thrust axis) may lead to some smearing if the proposed collective effects are strongest with respect to each individual jet direction. This is particularly an issue, when
a hard third jet is present. More careful studies would be needed.

{\textit{Winner-take-all jet recombination:}} As the first step of all 
proposed analyses is to rotate the lab frame to a new frame where
the jet direction represents the beam axis, the choice of the jet
axis (which is not unique) plays a crucial role. Besides the standard
``E-scheme'' recombination~\cite{Cacciari:2008gp} of jet reconstruction,
where the jet axis and the jet momentum are aligned at each stage 
of the recursion, the ``winner-take-all''
scheme~\cite{Bertolini:2013iqa,Larkoski:2014uqa} chooses the jet
axis to be align with the harder particle in a pair-wise recombination.
The motivation of the winner-take-all scheme is to minimize the impact
of soft radiation recoils to the initial parton direction.
It would be interesting to investigate how all observables would 
depend on different choices of jet axis.

{\textit{Lepton-lepton or lepton-proton/ion collisions:}} 
As all proposed studies take place within a single jet, 
they are in principle independent of the initial colliding beam
species, which can be protons, leptons or ions. Therefore,
these studies are highly relevant not only to the LHC but also 
all future high-energy colliders. In fact, $e^{+}e^{-}$ or $e^{-}p$ 
colliders may even provide a cleaner environment for studying
high-multiplicity jets, as the underlying event contribution is much smaller.

{\textit{Jets in heavy ion collisions:}} Finally, the study of 
parton energy loss in a QGP medium has been a main theme of research 
in $AA$ collisions. In our thought experiment, a parton propagating in the vacuum
or the QGP has no fundamental difference. In both cases, the parton
loses its energy by interacting with other partons along its passage.
The only difference is that in the QGP medium, surrounding partons 
are excited and thus have stronger color fields, which
lead to larger energy loss than interactions with vacuum 
chiral condensates. Doing the same analyses for jets in heavy ion 
collisions may provide insights to develop a unified approach to 
describing parton energy loss in confined and deconfined environments.

\section{SUMMARY}\label{sec:summary}  

Motivated by early surprises of thermal and collective 
phenomena in small system collisions,
we postulate that non-perturbative QCD evolution 
of a fragmenting parton in the vacuum will develop similar long-range collective 
effects to those of a multi-parton system, reminiscent of what is observed 
in high-energy hadronic or nuclear interactions with high-multiplicity 
final-state particles.
We propose searches for these properties of a parton
propagating in the vacuum using high-$\pT$ jets produced with 
large multiplicities in high-energy elementary collisions, e.g., at the LHC. 
A set of key observables are studied in detail
using the PYTHIA 8 Monte Carlo event generator, where no collective or QGP
effects are expected inside the jet. Experimental observation of the proposed effects (e.g., long-range collectivity or strangeness enhancement) in a single jet will offer a new view of non-perturbative QCD dynamics of multi-parton
systems at the smallest scales. On the other hand, absence of these effects may offer
new insights in to the role of quantum entanglement in the observed 
thermal behavior of particle production in high energy collisions.

\begin{acknowledgments}
The authors would like to thank Jamie Nagle, Zhoudunming Tu, Raju Venugopalan for useful discussions.
This work is in part supported by the Department of Energy 
grant number DE-SC0005131.
\end{acknowledgments}

\bibliography{Manuscript}

\begin{thebibliography}{85}%
\makeatletter
\providecommand \@ifxundefined [1]{%
 \@ifx{#1\undefined}
}%
\providecommand \@ifnum [1]{%
 \ifnum #1\expandafter \@firstoftwo
 \else \expandafter \@secondoftwo
 \fi
}%
\providecommand \@ifx [1]{%
 \ifx #1\expandafter \@firstoftwo
 \else \expandafter \@secondoftwo
 \fi
}%
\providecommand \natexlab [1]{#1}%
\providecommand \enquote  [1]{``#1''}%
\providecommand \bibnamefont  [1]{#1}%
\providecommand \bibfnamefont [1]{#1}%
\providecommand \citenamefont [1]{#1}%
\providecommand \href@noop [0]{\@secondoftwo}%
\providecommand \href [0]{\begingroup \@sanitize@url \@href}%
\providecommand \@href[1]{\@@startlink{#1}\@@href}%
\providecommand \@@href[1]{\endgroup#1\@@endlink}%
\providecommand \@sanitize@url [0]{\catcode `\\12\catcode `\$12\catcode
  `\&12\catcode `\#12\catcode `\^12\catcode `\_12\catcode `\%12\relax}%
\providecommand \@@startlink[1]{}%
\providecommand \@@endlink[0]{}%
\providecommand \url  [0]{\begingroup\@sanitize@url \@url }%
\providecommand \@url [1]{\endgroup\@href {#1}{\urlprefix }}%
\providecommand \urlprefix  [0]{URL }%
\providecommand \Eprint [0]{\href }%
\providecommand \doibase [0]{https://doi.org/}%
\providecommand \selectlanguage [0]{\@gobble}%
\providecommand \bibinfo  [0]{\@secondoftwo}%
\providecommand \bibfield  [0]{\@secondoftwo}%
\providecommand \translation [1]{[#1]}%
\providecommand \BibitemOpen [0]{}%
\providecommand \bibitemStop [0]{}%
\providecommand \bibitemNoStop [0]{.\EOS\space}%
\providecommand \EOS [0]{\spacefactor3000\relax}%
\providecommand \BibitemShut  [1]{\csname bibitem#1\endcsname}%
\let\auto@bib@innerbib\@empty
\bibitem [{\citenamefont {Gross}\ and\ \citenamefont
  {Wilczek}(1973)}]{Gross:1973id}%
  \BibitemOpen
  \bibfield  {author} {\bibinfo {author} {\bibfnamefont {D.~J.}\ \bibnamefont
  {Gross}}\ and\ \bibinfo {author} {\bibfnamefont {F.}~\bibnamefont
  {Wilczek}},\ }\bibfield  {title} {\bibinfo {title} {{Ultraviolet Behavior of
  Nonabelian Gauge Theories}},\ }\href
  {https://doi.org/10.1103/PhysRevLett.30.1343} {\bibfield  {journal} {\bibinfo
   {journal} {Phys.\ Rev.\ Lett.}\ }\textbf {\bibinfo {volume} {30}},\ \bibinfo
  {pages} {1343} (\bibinfo {year} {1973})}\BibitemShut {NoStop}%
\bibitem [{\citenamefont {Politzer}(1973)}]{Politzer:1973fx}%
  \BibitemOpen
  \bibfield  {author} {\bibinfo {author} {\bibfnamefont {H.}~\bibnamefont
  {Politzer}},\ }\bibfield  {title} {\bibinfo {title} {{Reliable Perturbative
  Results for Strong Interactions?}},\ }\href
  {https://doi.org/10.1103/PhysRevLett.30.1346} {\bibfield  {journal} {\bibinfo
   {journal} {Phys.\ Rev.\ Lett.}\ }\textbf {\bibinfo {volume} {30}},\ \bibinfo
  {pages} {1346} (\bibinfo {year} {1973})}\BibitemShut {NoStop}%
\bibitem [{\citenamefont {Brock}\ \emph {et~al.}(1995)\citenamefont {Brock}
  \emph {et~al.}}]{Brock:1993sz}%
  \BibitemOpen
  \bibfield  {author} {\bibinfo {author} {\bibfnamefont {R.}~\bibnamefont
  {Brock}} \emph {et~al.} (\bibinfo {collaboration} {CTEQ}),\ }\bibfield
  {title} {\bibinfo {title} {Handbook of perturbative qcd: Version 1.0},\
  }\href {https://doi.org/10.1103/RevModPhys.67.157} {\bibfield  {journal}
  {\bibinfo  {journal} {Rev. Mod. Phys.}\ }\textbf {\bibinfo {volume} {67}},\
  \bibinfo {pages} {157} (\bibinfo {year} {1995})}\BibitemShut {NoStop}%
\bibitem [{\citenamefont {Andersson}\ \emph {et~al.}(1983)\citenamefont
  {Andersson}, \citenamefont {Gustafson}, \citenamefont {Ingelman},\ and\
  \citenamefont {Sjostrand}}]{Andersson:1983ia}%
  \BibitemOpen
  \bibfield  {author} {\bibinfo {author} {\bibfnamefont {B.}~\bibnamefont
  {Andersson}}, \bibinfo {author} {\bibfnamefont {G.}~\bibnamefont
  {Gustafson}}, \bibinfo {author} {\bibfnamefont {G.}~\bibnamefont
  {Ingelman}},\ and\ \bibinfo {author} {\bibfnamefont {T.}~\bibnamefont
  {Sjostrand}},\ }\bibfield  {title} {\bibinfo {title} {{Parton Fragmentation
  and String Dynamics}},\ }\href {https://doi.org/10.1016/0370-1573(83)90080-7}
  {\bibfield  {journal} {\bibinfo  {journal} {Phys. Rept.}\ }\textbf {\bibinfo
  {volume} {97}},\ \bibinfo {pages} {31} (\bibinfo {year} {1983})}\BibitemShut
  {NoStop}%
\bibitem [{\citenamefont {Marchesini}\ \emph {et~al.}(1992)\citenamefont
  {Marchesini}, \citenamefont {Webber}, \citenamefont {Abbiendi}, \citenamefont
  {Knowles}, \citenamefont {Seymour},\ and\ \citenamefont
  {Stanco}}]{Marchesini:1991ch}%
  \BibitemOpen
  \bibfield  {author} {\bibinfo {author} {\bibfnamefont {G.}~\bibnamefont
  {Marchesini}}, \bibinfo {author} {\bibfnamefont {B.}~\bibnamefont {Webber}},
  \bibinfo {author} {\bibfnamefont {G.}~\bibnamefont {Abbiendi}}, \bibinfo
  {author} {\bibfnamefont {I.}~\bibnamefont {Knowles}}, \bibinfo {author}
  {\bibfnamefont {M.}~\bibnamefont {Seymour}},\ and\ \bibinfo {author}
  {\bibfnamefont {L.}~\bibnamefont {Stanco}},\ }\bibfield  {title} {\bibinfo
  {title} {{HERWIG: A Monte Carlo event generator for simulating hadron
  emission reactions with interfering gluons. Version 5.1 - April 1991}},\
  }\href {https://doi.org/10.1016/0010-4655(92)90055-4} {\bibfield  {journal}
  {\bibinfo  {journal} {Comput. Phys. Commun.}\ }\textbf {\bibinfo {volume}
  {67}},\ \bibinfo {pages} {465} (\bibinfo {year} {1992})}\BibitemShut
  {NoStop}%
\bibitem [{\citenamefont {Kogler}\ \emph {et~al.}(2019)\citenamefont {Kogler}
  \emph {et~al.}}]{Asquith:2018igt}%
  \BibitemOpen
  \bibfield  {author} {\bibinfo {author} {\bibfnamefont {R.}~\bibnamefont
  {Kogler}} \emph {et~al.},\ }\bibfield  {title} {\bibinfo {title} {{Jet
  Substructure at the Large Hadron Collider: Experimental Review}},\ }\href
  {https://doi.org/10.1103/RevModPhys.91.045003} {\bibfield  {journal}
  {\bibinfo  {journal} {Rev. Mod. Phys.}\ }\textbf {\bibinfo {volume} {91}},\
  \bibinfo {pages} {045003} (\bibinfo {year} {2019})},\ \Eprint
  {https://arxiv.org/abs/1803.06991} {arXiv:1803.06991 [hep-ex]} \BibitemShut
  {NoStop}%
\bibitem [{\citenamefont {Karsch}(1995)}]{Karsch:1995sy}%
  \BibitemOpen
  \bibfield  {author} {\bibinfo {author} {\bibfnamefont {F.}~\bibnamefont
  {Karsch}},\ }\bibfield  {title} {\bibinfo {title} {{The Phase transition to
  the quark gluon plasma: Recent results from lattice calculations}},\ }\href
  {https://doi.org/10.1016/0375-9474(95)00248-Y} {\bibfield  {journal}
  {\bibinfo  {journal} {Nucl. Phys. A}\ }\textbf {\bibinfo {volume} {590}},\
  \bibinfo {pages} {367C} (\bibinfo {year} {1995})},\ \Eprint
  {https://arxiv.org/abs/hep-lat/9503010} {arXiv:hep-lat/9503010 [hep-lat]}
  \BibitemShut {NoStop}%
\bibitem [{\citenamefont {Karsch}(2002)}]{Karsch:2001cy}%
  \BibitemOpen
  \bibfield  {author} {\bibinfo {author} {\bibfnamefont {F.}~\bibnamefont
  {Karsch}},\ }\bibfield  {title} {\bibinfo {title} {{Lattice QCD at high
  temperature and density}},\ }\href@noop {} {\bibfield  {journal} {\bibinfo
  {journal} {Lect. Notes Phys.}\ }\textbf {\bibinfo {volume} {583}},\ \bibinfo
  {pages} {209} (\bibinfo {year} {2002})},\ \Eprint
  {https://arxiv.org/abs/hep-lat/0106019} {arXiv:hep-lat/0106019 [hep-lat]}
  \BibitemShut {NoStop}%
\bibitem [{\citenamefont {Bazavov}\ \emph {et~al.}(2019)\citenamefont {Bazavov}
  \emph {et~al.}}]{Bazavov:2018mes}%
  \BibitemOpen
  \bibfield  {author} {\bibinfo {author} {\bibfnamefont {A.}~\bibnamefont
  {Bazavov}} \emph {et~al.} (\bibinfo {collaboration} {HotQCD}),\ }\bibfield
  {title} {\bibinfo {title} {{Chiral crossover in QCD at zero and non-zero
  chemical potentials}},\ }\href
  {https://doi.org/10.1016/j.physletb.2019.05.013} {\bibfield  {journal}
  {\bibinfo  {journal} {Phys. Lett. B}\ }\textbf {\bibinfo {volume} {795}},\
  \bibinfo {pages} {15} (\bibinfo {year} {2019})},\ \Eprint
  {https://arxiv.org/abs/1812.08235} {arXiv:1812.08235 [hep-lat]} \BibitemShut
  {NoStop}%
\bibitem [{CER(2020)}]{CERN-SPS}%
  \BibitemOpen
  \bibfield  {title} {\bibinfo {title} {{CERN Press Release Feb. 10}}}
  (\bibinfo {year} {2020})\BibitemShut {NoStop}%
\bibitem [{\citenamefont {Arsene}\ \emph {et~al.}(2005)\citenamefont {Arsene}
  \emph {et~al.}}]{BRAHMS}%
  \BibitemOpen
  \bibfield  {author} {\bibinfo {author} {\bibfnamefont {I.}~\bibnamefont
  {Arsene}} \emph {et~al.} (\bibinfo {collaboration} {BRAHMS}),\ }\bibfield
  {title} {\bibinfo {title} {{Quark gluon plasma and color glass condensate at
  RHIC? The Perspective from the BRAHMS experiment}},\ }\href
  {https://doi.org/10.1016/j.nuclphysa.2005.02.130} {\bibfield  {journal}
  {\bibinfo  {journal} {Nucl. Phys. A}\ }\textbf {\bibinfo {volume} {757}},\
  \bibinfo {pages} {1} (\bibinfo {year} {2005})},\ \Eprint
  {https://arxiv.org/abs/nucl-ex/0410020} {arXiv:nucl-ex/0410020 [nucl-ex]}
  \BibitemShut {NoStop}%
\bibitem [{\citenamefont {Adcox}\ \emph {et~al.}(2005)\citenamefont {Adcox}
  \emph {et~al.}}]{PHENIX}%
  \BibitemOpen
  \bibfield  {author} {\bibinfo {author} {\bibfnamefont {K.}~\bibnamefont
  {Adcox}} \emph {et~al.} (\bibinfo {collaboration} {PHENIX}),\ }\bibfield
  {title} {\bibinfo {title} {{Formation of dense partonic matter in
  relativistic nucleus-nucleus collisions at RHIC: Experimental evaluation by
  the PHENIX collaboration}},\ }\href
  {https://doi.org/10.1016/j.nuclphysa.2005.03.086} {\bibfield  {journal}
  {\bibinfo  {journal} {Nucl. Phys. A}\ }\textbf {\bibinfo {volume} {757}},\
  \bibinfo {pages} {184} (\bibinfo {year} {2005})},\ \Eprint
  {https://arxiv.org/abs/nucl-ex/0410003} {arXiv:nucl-ex/0410003 [nucl-ex]}
  \BibitemShut {NoStop}%
\bibitem [{\citenamefont {Back}\ \emph {et~al.}(2005)\citenamefont {Back} \emph
  {et~al.}}]{PHOBOS}%
  \BibitemOpen
  \bibfield  {author} {\bibinfo {author} {\bibfnamefont {B.~B.}\ \bibnamefont
  {Back}} \emph {et~al.} (\bibinfo {collaboration} {PHOBOS}),\ }\bibfield
  {title} {\bibinfo {title} {{The PHOBOS perspective on discoveries at RHIC}},\
  }\href {https://doi.org/10.1016/j.nuclphysa.2005.03.084} {\bibfield
  {journal} {\bibinfo  {journal} {Nucl. Phys. A}\ }\textbf {\bibinfo {volume}
  {757}},\ \bibinfo {pages} {28} (\bibinfo {year} {2005})},\ \Eprint
  {https://arxiv.org/abs/nucl-ex/0410022} {arXiv:nucl-ex/0410022 [nucl-ex]}
  \BibitemShut {NoStop}%
\bibitem [{\citenamefont {Adams}\ \emph
  {et~al.}(2005{\natexlab{a}})\citenamefont {Adams} \emph {et~al.}}]{STAR}%
  \BibitemOpen
  \bibfield  {author} {\bibinfo {author} {\bibfnamefont {J.}~\bibnamefont
  {Adams}} \emph {et~al.} (\bibinfo {collaboration} {STAR}),\ }\bibfield
  {title} {\bibinfo {title} {{Experimental and theoretical challenges in the
  search for the quark gluon plasma: The STAR Collaboration's critical
  assessment of the evidence from RHIC collisions}},\ }\href
  {https://doi.org/10.1016/j.nuclphysa.2005.03.085} {\bibfield  {journal}
  {\bibinfo  {journal} {Nucl. Phys. A}\ }\textbf {\bibinfo {volume} {757}},\
  \bibinfo {pages} {102} (\bibinfo {year} {2005}{\natexlab{a}})},\ \Eprint
  {https://arxiv.org/abs/nucl-ex/0501009} {arXiv:nucl-ex/0501009 [nucl-ex]}
  \BibitemShut {NoStop}%
\bibitem [{\citenamefont {Muller}\ \emph {et~al.}(2012)\citenamefont {Muller},
  \citenamefont {Schukraft},\ and\ \citenamefont {Wyslouch}}]{Muller:2012zq}%
  \BibitemOpen
  \bibfield  {author} {\bibinfo {author} {\bibfnamefont {B.}~\bibnamefont
  {Muller}}, \bibinfo {author} {\bibfnamefont {J.}~\bibnamefont {Schukraft}},\
  and\ \bibinfo {author} {\bibfnamefont {B.}~\bibnamefont {Wyslouch}},\
  }\bibfield  {title} {\bibinfo {title} {{First Results from Pb+Pb collisions
  at the LHC}},\ }\href {https://doi.org/10.1146/annurev-nucl-102711-094910}
  {\bibfield  {journal} {\bibinfo  {journal} {Ann. Rev. Nucl. Part. Sci.}\
  }\textbf {\bibinfo {volume} {62}},\ \bibinfo {pages} {361} (\bibinfo {year}
  {2012})},\ \Eprint {https://arxiv.org/abs/1202.3233} {arXiv:1202.3233
  [hep-ex]} \BibitemShut {NoStop}%
\bibitem [{\citenamefont {Adams}\ \emph
  {et~al.}(2005{\natexlab{b}})\citenamefont {Adams} \emph
  {et~al.}}]{Adams:2005ph}%
  \BibitemOpen
  \bibfield  {author} {\bibinfo {author} {\bibfnamefont {J.}~\bibnamefont
  {Adams}} \emph {et~al.} (\bibinfo {collaboration} {STAR}),\ }\bibfield
  {title} {\bibinfo {title} {{Distributions of charged hadrons associated with
  high transverse momentum particles in pp and Au + Au collisions at \rootsNN\
  = 200\GeV}},\ }\href {https://doi.org/10.1103/PhysRevLett.95.152301}
  {\bibfield  {journal} {\bibinfo  {journal} {Phys. Rev. Lett.}\ }\textbf
  {\bibinfo {volume} {95}},\ \bibinfo {pages} {152301} (\bibinfo {year}
  {2005}{\natexlab{b}})},\ \Eprint {https://arxiv.org/abs/nucl-ex/0501016}
  {nucl-ex/0501016} \BibitemShut {NoStop}%
\bibitem [{\citenamefont {Alver}\ \emph
  {et~al.}(2010{\natexlab{a}})\citenamefont {Alver} \emph
  {et~al.}}]{Alver:2008gk}%
  \BibitemOpen
  \bibfield  {author} {\bibinfo {author} {\bibfnamefont {B.}~\bibnamefont
  {Alver}} \emph {et~al.} (\bibinfo {collaboration} {PHOBOS}),\ }\bibfield
  {title} {\bibinfo {title} {{System size dependence of cluster properties from
  two- particle angular correlations in Cu+Cu and Au+Au collisions at \rootsNN\
  = 200\GeV}},\ }\href {https://doi.org/10.1103/PhysRevC.81.024904} {\bibfield
  {journal} {\bibinfo  {journal} {Phys. Rev. C}\ }\textbf {\bibinfo {volume}
  {81}},\ \bibinfo {pages} {024904} (\bibinfo {year} {2010}{\natexlab{a}})},\
  \Eprint {https://arxiv.org/abs/0812.1172} {arXiv:0812.1172 [nucl-ex]}
  \BibitemShut {NoStop}%
\bibitem [{\citenamefont {Abelev}\ \emph {et~al.}(2009)\citenamefont {Abelev}
  \emph {et~al.}}]{Abelev:2009af}%
  \BibitemOpen
  \bibfield  {author} {\bibinfo {author} {\bibfnamefont {B.}~\bibnamefont
  {Abelev}} \emph {et~al.} (\bibinfo {collaboration} {STAR}),\ }\bibfield
  {title} {\bibinfo {title} {{Long range rapidity correlations and jet
  production in high energy nuclear collisions}},\ }\href
  {https://doi.org/10.1103/PhysRevC.80.064912} {\bibfield  {journal} {\bibinfo
  {journal} {Phys. Rev. C}\ }\textbf {\bibinfo {volume} {80}},\ \bibinfo
  {pages} {064912} (\bibinfo {year} {2009})},\ \Eprint
  {https://arxiv.org/abs/0909.0191} {arXiv:0909.0191 [nucl-ex]} \BibitemShut
  {NoStop}%
\bibitem [{\citenamefont {Alver}\ \emph
  {et~al.}(2010{\natexlab{b}})\citenamefont {Alver} \emph
  {et~al.}}]{Alver:2009id}%
  \BibitemOpen
  \bibfield  {author} {\bibinfo {author} {\bibfnamefont {B.}~\bibnamefont
  {Alver}} \emph {et~al.} (\bibinfo {collaboration} {PHOBOS}),\ }\bibfield
  {title} {\bibinfo {title} {{High transverse momentum triggered correlations
  over a large pseudorapidity acceptance in Au+Au collisions at \rootsNN\ =
  200\GeV}},\ }\href {https://doi.org/10.1103/PhysRevLett.104.062301}
  {\bibfield  {journal} {\bibinfo  {journal} {Phys. Rev. Lett.}\ }\textbf
  {\bibinfo {volume} {104}},\ \bibinfo {pages} {062301} (\bibinfo {year}
  {2010}{\natexlab{b}})},\ \Eprint {https://arxiv.org/abs/0903.2811}
  {arXiv:0903.2811 [nucl-ex]} \BibitemShut {NoStop}%
\bibitem [{\citenamefont {Chatrchyan}\ \emph {et~al.}(2011)\citenamefont
  {Chatrchyan} \emph {et~al.}}]{Chatrchyan:2011eka}%
  \BibitemOpen
  \bibfield  {author} {\bibinfo {author} {\bibfnamefont {S.}~\bibnamefont
  {Chatrchyan}} \emph {et~al.} (\bibinfo {collaboration} {CMS}),\ }\bibfield
  {title} {\bibinfo {title} {{Long-range and short-range dihadron angular
  correlations in central PbPb collisions at a nucleon-nucleon center of mass
  energy of 2.76 TeV}},\ }\href {https://doi.org/10.1007/JHEP07(2011)076}
  {\bibfield  {journal} {\bibinfo  {journal} {JHEP}\ }\textbf {\bibinfo
  {volume} {07}},\ \bibinfo {pages} {076}},\ \Eprint
  {https://arxiv.org/abs/1105.2438} {arXiv:1105.2438 [nucl-ex]} \BibitemShut
  {NoStop}%
\bibitem [{\citenamefont {Chatrchyan}\ \emph {et~al.}(2012)\citenamefont
  {Chatrchyan} \emph {et~al.}}]{Chatrchyan:2012wg}%
  \BibitemOpen
  \bibfield  {author} {\bibinfo {author} {\bibfnamefont {S.}~\bibnamefont
  {Chatrchyan}} \emph {et~al.} (\bibinfo {collaboration} {CMS}),\ }\bibfield
  {title} {\bibinfo {title} {{Centrality dependence of dihadron correlations
  and azimuthal anisotropy harmonics in PbPb collisions at \rootsNN\ = 2.76
  TeV}},\ }\href {https://doi.org/10.1140/epjc/s10052-012-2012-3} {\bibfield
  {journal} {\bibinfo  {journal} {Eur. Phys. J. C}\ }\textbf {\bibinfo {volume}
  {72}},\ \bibinfo {pages} {2012} (\bibinfo {year} {2012})},\ \Eprint
  {https://arxiv.org/abs/1201.3158} {arXiv:1201.3158 [nucl-ex]} \BibitemShut
  {NoStop}%
\bibitem [{\citenamefont {Aamodt}\ \emph {et~al.}(2010)\citenamefont {Aamodt}
  \emph {et~al.}}]{Aamodt:2010pa}%
  \BibitemOpen
  \bibfield  {author} {\bibinfo {author} {\bibfnamefont {K.}~\bibnamefont
  {Aamodt}} \emph {et~al.} (\bibinfo {collaboration} {ALICE}),\ }\bibfield
  {title} {\bibinfo {title} {{Elliptic flow of charged particles in Pb-Pb
  collisions at 2.76 TeV}},\ }\href
  {https://doi.org/10.1103/PhysRevLett.105.252302} {\bibfield  {journal}
  {\bibinfo  {journal} {Phys. Rev. Lett.}\ }\textbf {\bibinfo {volume} {105}},\
  \bibinfo {pages} {252302} (\bibinfo {year} {2010})},\ \Eprint
  {https://arxiv.org/abs/1011.3914} {arXiv:1011.3914 [nucl-ex]} \BibitemShut
  {NoStop}%
\bibitem [{\citenamefont {Aad}\ \emph {et~al.}(2012)\citenamefont {Aad} \emph
  {et~al.}}]{ATLAS:2012at}%
  \BibitemOpen
  \bibfield  {author} {\bibinfo {author} {\bibfnamefont {G.}~\bibnamefont
  {Aad}} \emph {et~al.} (\bibinfo {collaboration} {ATLAS}),\ }\bibfield
  {title} {\bibinfo {title} {{Measurement of the azimuthal anisotropy for
  charged particle production in \rootsNN\ = 2.76\TeV lead-lead collisions with
  the ATLAS detector}},\ }\href {https://doi.org/10.1103/PhysRevC.86.014907}
  {\bibfield  {journal} {\bibinfo  {journal} {Phys. Rev. C}\ }\textbf {\bibinfo
  {volume} {86}},\ \bibinfo {pages} {014907} (\bibinfo {year} {2012})},\
  \Eprint {https://arxiv.org/abs/1203.3087} {arXiv:1203.3087 [hep-ex]}
  \BibitemShut {NoStop}%
\bibitem [{\citenamefont {Chatrchyan}\ \emph {et~al.}(2014)\citenamefont
  {Chatrchyan} \emph {et~al.}}]{CMS:2013bza}%
  \BibitemOpen
  \bibfield  {author} {\bibinfo {author} {\bibfnamefont {S.}~\bibnamefont
  {Chatrchyan}} \emph {et~al.} (\bibinfo {collaboration} {CMS}),\ }\bibfield
  {title} {\bibinfo {title} {{Studies of azimuthal dihadron correlations in
  ultra-central PbPb collisions at \rootsNN\ = 2.76\TeV}},\ }\href
  {https://doi.org/10.1007/JHEP02(2014)088} {\bibfield  {journal} {\bibinfo
  {journal} {JHEP}\ }\textbf {\bibinfo {volume} {02}},\ \bibinfo {pages}
  {088}},\ \Eprint {https://arxiv.org/abs/1312.1845} {arXiv:1312.1845
  [nucl-ex]} \BibitemShut {NoStop}%
\bibitem [{\citenamefont {Ollitrault}(1992)}]{Ollitrault:1992bk}%
  \BibitemOpen
  \bibfield  {author} {\bibinfo {author} {\bibfnamefont {J.-Y.}\ \bibnamefont
  {Ollitrault}},\ }\bibfield  {title} {\bibinfo {title} {{Anisotropy as a
  signature of transverse collective flow}},\ }\href
  {https://doi.org/10.1103/PhysRevD.46.229} {\bibfield  {journal} {\bibinfo
  {journal} {Phys. Rev. D}\ }\textbf {\bibinfo {volume} {46}},\ \bibinfo
  {pages} {229} (\bibinfo {year} {1992})}\BibitemShut {NoStop}%
\bibitem [{\citenamefont {Teaney}(2003)}]{Teaney:2003kp}%
  \BibitemOpen
  \bibfield  {author} {\bibinfo {author} {\bibfnamefont {D.}~\bibnamefont
  {Teaney}},\ }\bibfield  {title} {\bibinfo {title} {{The Effects of viscosity
  on spectra, elliptic flow, and HBT radii}},\ }\href
  {https://doi.org/10.1103/PhysRevC.68.034913} {\bibfield  {journal} {\bibinfo
  {journal} {Phys. Rev. C}\ }\textbf {\bibinfo {volume} {68}},\ \bibinfo
  {pages} {034913} (\bibinfo {year} {2003})},\ \Eprint
  {https://arxiv.org/abs/nucl-th/0301099} {arXiv:nucl-th/0301099} \BibitemShut
  {NoStop}%
\bibitem [{\citenamefont {Romatschke}\ and\ \citenamefont
  {Romatschke}(2007)}]{Romatschke:2007mq}%
  \BibitemOpen
  \bibfield  {author} {\bibinfo {author} {\bibfnamefont {P.}~\bibnamefont
  {Romatschke}}\ and\ \bibinfo {author} {\bibfnamefont {U.}~\bibnamefont
  {Romatschke}},\ }\bibfield  {title} {\bibinfo {title} {{Viscosity Information
  from Relativistic Nuclear Collisions: How Perfect is the Fluid Observed at
  RHIC?}},\ }\href {https://doi.org/10.1103/PhysRevLett.99.172301} {\bibfield
  {journal} {\bibinfo  {journal} {Phys. Rev. Lett.}\ }\textbf {\bibinfo
  {volume} {99}},\ \bibinfo {pages} {172301} (\bibinfo {year} {2007})},\
  \Eprint {https://arxiv.org/abs/0706.1522} {arXiv:0706.1522 [nucl-th]}
  \BibitemShut {NoStop}%
\bibitem [{\citenamefont {Heinz}\ and\ \citenamefont
  {Snellings}(2013)}]{Heinz:2013th}%
  \BibitemOpen
  \bibfield  {author} {\bibinfo {author} {\bibfnamefont {U.}~\bibnamefont
  {Heinz}}\ and\ \bibinfo {author} {\bibfnamefont {R.}~\bibnamefont
  {Snellings}},\ }\bibfield  {title} {\bibinfo {title} {{Collective flow and
  viscosity in relativistic heavy-ion collisions}},\ }\href
  {https://doi.org/10.1146/annurev-nucl-102212-170540} {\bibfield  {journal}
  {\bibinfo  {journal} {Ann. Rev. Nucl. Part. Sci.}\ }\textbf {\bibinfo
  {volume} {63}},\ \bibinfo {pages} {123} (\bibinfo {year} {2013})},\ \Eprint
  {https://arxiv.org/abs/1301.2826} {arXiv:1301.2826 [nucl-th]} \BibitemShut
  {NoStop}%
\bibitem [{\citenamefont {Gale}\ \emph {et~al.}(2013)\citenamefont {Gale},
  \citenamefont {Jeon},\ and\ \citenamefont {Schenke}}]{Gale:2013da}%
  \BibitemOpen
  \bibfield  {author} {\bibinfo {author} {\bibfnamefont {C.}~\bibnamefont
  {Gale}}, \bibinfo {author} {\bibfnamefont {S.}~\bibnamefont {Jeon}},\ and\
  \bibinfo {author} {\bibfnamefont {B.}~\bibnamefont {Schenke}},\ }\bibfield
  {title} {\bibinfo {title} {{Hydrodynamic Modeling of Heavy-Ion Collisions}},\
  }\href {https://doi.org/10.1142/S0217751X13400113} {\bibfield  {journal}
  {\bibinfo  {journal} {Int. J. Mod. Phys. A}\ }\textbf {\bibinfo {volume}
  {28}},\ \bibinfo {pages} {1340011} (\bibinfo {year} {2013})},\ \Eprint
  {https://arxiv.org/abs/1301.5893} {arXiv:1301.5893 [nucl-th]} \BibitemShut
  {NoStop}%
\bibitem [{\citenamefont {Khachatryan}\ \emph {et~al.}(2010)\citenamefont
  {Khachatryan} \emph {et~al.}}]{Khachatryan:2010gv}%
  \BibitemOpen
  \bibfield  {author} {\bibinfo {author} {\bibfnamefont {V.}~\bibnamefont
  {Khachatryan}} \emph {et~al.} (\bibinfo {collaboration} {CMS}),\ }\bibfield
  {title} {\bibinfo {title} {{Observation of Long-Range Near-Side Angular
  Correlations in Proton-Proton Collisions at the LHC}},\ }\href
  {https://doi.org/10.1007/JHEP09(2010)091} {\bibfield  {journal} {\bibinfo
  {journal} {JHEP}\ }\textbf {\bibinfo {volume} {09}},\ \bibinfo {pages}
  {091}},\ \Eprint {https://arxiv.org/abs/1009.4122} {arXiv:1009.4122 [hep-ex]}
  \BibitemShut {NoStop}%
\bibitem [{\citenamefont {Aad}\ \emph {et~al.}(2016{\natexlab{a}})\citenamefont
  {Aad} \emph {et~al.}}]{Aad:2015gqa}%
  \BibitemOpen
  \bibfield  {author} {\bibinfo {author} {\bibfnamefont {G.}~\bibnamefont
  {Aad}} \emph {et~al.} (\bibinfo {collaboration} {ATLAS}),\ }\bibfield
  {title} {\bibinfo {title} {{Observation of Long-Range Elliptic Azimuthal
  Anisotropies in $\sqrt{s}=$ 13 and 2.76 TeV $pp$ Collisions with the ATLAS
  Detector}},\ }\href {https://doi.org/10.1103/PhysRevLett.116.172301}
  {\bibfield  {journal} {\bibinfo  {journal} {Phys. Rev. Lett.}\ }\textbf
  {\bibinfo {volume} {116}},\ \bibinfo {pages} {172301} (\bibinfo {year}
  {2016}{\natexlab{a}})},\ \Eprint {https://arxiv.org/abs/1509.04776}
  {arXiv:1509.04776 [hep-ex]} \BibitemShut {NoStop}%
\bibitem [{\citenamefont {Khachatryan}\ \emph {et~al.}(2016)\citenamefont
  {Khachatryan} \emph {et~al.}}]{Khachatryan:2015lva}%
  \BibitemOpen
  \bibfield  {author} {\bibinfo {author} {\bibfnamefont {V.}~\bibnamefont
  {Khachatryan}} \emph {et~al.} (\bibinfo {collaboration} {CMS}),\ }\bibfield
  {title} {\bibinfo {title} {{Measurement of long-range near-side two-particle
  angular correlations in pp collisions at $\sqrt s =$13 TeV}},\ }\href
  {https://doi.org/10.1103/PhysRevLett.116.172302} {\bibfield  {journal}
  {\bibinfo  {journal} {Phys. Rev. Lett.}\ }\textbf {\bibinfo {volume} {116}},\
  \bibinfo {pages} {172302} (\bibinfo {year} {2016})},\ \Eprint
  {https://arxiv.org/abs/1510.03068} {arXiv:1510.03068 [nucl-ex]} \BibitemShut
  {NoStop}%
\bibitem [{\citenamefont {Khachatryan}\ \emph
  {et~al.}(2017{\natexlab{a}})\citenamefont {Khachatryan} \emph
  {et~al.}}]{Khachatryan:2016txc}%
  \BibitemOpen
  \bibfield  {author} {\bibinfo {author} {\bibfnamefont {V.}~\bibnamefont
  {Khachatryan}} \emph {et~al.} (\bibinfo {collaboration} {CMS}),\ }\bibfield
  {title} {\bibinfo {title} {{Evidence for collectivity in pp collisions at the
  LHC}},\ }\href {https://doi.org/10.1016/j.physletb.2016.12.009} {\bibfield
  {journal} {\bibinfo  {journal} {Phys. Lett. B}\ }\textbf {\bibinfo {volume}
  {765}},\ \bibinfo {pages} {193} (\bibinfo {year} {2017}{\natexlab{a}})},\
  \Eprint {https://arxiv.org/abs/1606.06198} {arXiv:1606.06198 [nucl-ex]}
  \BibitemShut {NoStop}%
\bibitem [{\citenamefont {Aad}\ \emph {et~al.}(2020)\citenamefont {Aad} \emph
  {et~al.}}]{Aad:2019aol}%
  \BibitemOpen
  \bibfield  {author} {\bibinfo {author} {\bibfnamefont {G.}~\bibnamefont
  {Aad}} \emph {et~al.} (\bibinfo {collaboration} {ATLAS}),\ }\bibfield
  {title} {\bibinfo {title} {{Measurement of azimuthal anisotropy of muons from
  charm and bottom hadrons in $pp$ collisions at $\sqrt{s}=13$ TeV with the
  ATLAS detector}},\ }\href {https://doi.org/10.1103/PhysRevLett.124.082301}
  {\bibfield  {journal} {\bibinfo  {journal} {Phys. Rev. Lett.}\ }\textbf
  {\bibinfo {volume} {124}},\ \bibinfo {pages} {082301} (\bibinfo {year}
  {2020})},\ \Eprint {https://arxiv.org/abs/1909.01650} {arXiv:1909.01650
  [nucl-ex]} \BibitemShut {NoStop}%
\bibitem [{\citenamefont {Li}(2012)}]{Li:2012hc}%
  \BibitemOpen
  \bibfield  {author} {\bibinfo {author} {\bibfnamefont {W.}~\bibnamefont
  {Li}},\ }\bibfield  {title} {\bibinfo {title} {{Observation of a 'Ridge'
  correlation structure in high multiplicity proton-proton collisions: A brief
  review}},\ }\href {https://doi.org/10.1142/S0217732312300182} {\bibfield
  {journal} {\bibinfo  {journal} {Mod. Phys. Lett. A}\ }\textbf {\bibinfo
  {volume} {27}},\ \bibinfo {pages} {1230018} (\bibinfo {year} {2012})},\
  \Eprint {https://arxiv.org/abs/1206.0148} {arXiv:1206.0148 [nucl-ex]}
  \BibitemShut {NoStop}%
\bibitem [{\citenamefont {Chatrchyan}\ \emph {et~al.}(2013)\citenamefont
  {Chatrchyan} \emph {et~al.}}]{CMS:2012qk}%
  \BibitemOpen
  \bibfield  {author} {\bibinfo {author} {\bibfnamefont {S.}~\bibnamefont
  {Chatrchyan}} \emph {et~al.} (\bibinfo {collaboration} {CMS}),\ }\bibfield
  {title} {\bibinfo {title} {{Observation of long-range near-side angular
  correlations in proton-lead collisions at the LHC}},\ }\href
  {https://doi.org/10.1016/j.physletb.2012.11.025} {\bibfield  {journal}
  {\bibinfo  {journal} {Phys. Lett. B}\ }\textbf {\bibinfo {volume} {718}},\
  \bibinfo {pages} {795} (\bibinfo {year} {2013})},\ \Eprint
  {https://arxiv.org/abs/1210.5482} {arXiv:1210.5482 [nucl-ex]} \BibitemShut
  {NoStop}%
\bibitem [{\citenamefont {Abelev}\ \emph
  {et~al.}(2013{\natexlab{a}})\citenamefont {Abelev} \emph
  {et~al.}}]{alice:2012qe}%
  \BibitemOpen
  \bibfield  {author} {\bibinfo {author} {\bibfnamefont {B.}~\bibnamefont
  {Abelev}} \emph {et~al.} (\bibinfo {collaboration} {ALICE}),\ }\bibfield
  {title} {\bibinfo {title} {{Long-range angular correlations on the near and
  away side in pPb collisions at $\sqrt{s_{\text{NN}}}$ = 5.02 TeV }},\ }\href
  {https://doi.org/10.1016/j.physletb.2013.01.012} {\bibfield  {journal}
  {\bibinfo  {journal} {Phys. Lett. B}\ }\textbf {\bibinfo {volume} {719}},\
  \bibinfo {pages} {29} (\bibinfo {year} {2013}{\natexlab{a}})},\ \Eprint
  {https://arxiv.org/abs/1212.2001} {arXiv:1212.2001 [nucl-ex]} \BibitemShut
  {NoStop}%
\bibitem [{\citenamefont {Aad}\ \emph {et~al.}(2013)\citenamefont {Aad} \emph
  {et~al.}}]{Aad:2012gla}%
  \BibitemOpen
  \bibfield  {author} {\bibinfo {author} {\bibfnamefont {G.}~\bibnamefont
  {Aad}} \emph {et~al.} (\bibinfo {collaboration} {ATLAS}),\ }\bibfield
  {title} {\bibinfo {title} {Observation of associated near-side and away-side
  long-range correlations in {$\sqrt{s_{NN}} = 5.02 TeV$} proton-lead
  collisions with the {ATLAS} detector},\ }\href
  {https://doi.org/10.1103/PhysRevLett.110.182302} {\bibfield  {journal}
  {\bibinfo  {journal} {Phys. Rev. Lett.}\ }\textbf {\bibinfo {volume} {110}},\
  \bibinfo {pages} {182302} (\bibinfo {year} {2013})},\ \Eprint
  {https://arxiv.org/abs/1212.5198} {arXiv:1212.5198 [hep-ex]} \BibitemShut
  {NoStop}%
\bibitem [{\citenamefont {Aaij}\ \emph {et~al.}(2016)\citenamefont {Aaij} \emph
  {et~al.}}]{Aaij:2015qcq}%
  \BibitemOpen
  \bibfield  {author} {\bibinfo {author} {\bibfnamefont {R.}~\bibnamefont
  {Aaij}} \emph {et~al.} (\bibinfo {collaboration} {LHCb}),\ }\bibfield
  {title} {\bibinfo {title} {{Measurements of long-range near-side angular
  correlations in $\sqrt{s_{\text{NN}}}=5$ TeV proton-lead collisions in the
  forward region}},\ }\href {https://doi.org/10.1016/j.physletb.2016.09.064}
  {\bibfield  {journal} {\bibinfo  {journal} {Phys. Lett. B}\ }\textbf
  {\bibinfo {volume} {762}},\ \bibinfo {pages} {473} (\bibinfo {year}
  {2016})},\ \Eprint {https://arxiv.org/abs/1512.00439} {arXiv:1512.00439
  [nucl-ex]} \BibitemShut {NoStop}%
\bibitem [{\citenamefont {Abelev}\ \emph
  {et~al.}(2013{\natexlab{b}})\citenamefont {Abelev} \emph
  {et~al.}}]{ABELEV:2013wsa}%
  \BibitemOpen
  \bibfield  {author} {\bibinfo {author} {\bibfnamefont {B.~B.}\ \bibnamefont
  {Abelev}} \emph {et~al.} (\bibinfo {collaboration} {ALICE}),\ }\bibfield
  {title} {\bibinfo {title} {{Long-range angular correlations of pi, K and p in
  p--Pb collisions at $\sqrt{s_{\text{NN}}}$ = 5.02 TeV}},\ }\href
  {https://doi.org/10.1016/j.physletb.2013.08.024} {\bibfield  {journal}
  {\bibinfo  {journal} {Phys. Lett. B}\ }\textbf {\bibinfo {volume} {726}},\
  \bibinfo {pages} {164} (\bibinfo {year} {2013}{\natexlab{b}})},\ \Eprint
  {https://arxiv.org/abs/1307.3237} {arXiv:1307.3237 [nucl-ex]} \BibitemShut
  {NoStop}%
\bibitem [{\citenamefont {Khachatryan}\ \emph
  {et~al.}(2015{\natexlab{a}})\citenamefont {Khachatryan} \emph
  {et~al.}}]{Khachatryan:2014jra}%
  \BibitemOpen
  \bibfield  {author} {\bibinfo {author} {\bibfnamefont {V.}~\bibnamefont
  {Khachatryan}} \emph {et~al.} (\bibinfo {collaboration} {CMS}),\ }\bibfield
  {title} {\bibinfo {title} {{Long-range two-particle correlations of strange
  hadrons with charged particles in pPb and PbPb collisions at LHC energies}},\
  }\href {https://doi.org/10.1016/j.physletb.2015.01.034} {\bibfield  {journal}
  {\bibinfo  {journal} {Phys. Lett. B}\ }\textbf {\bibinfo {volume} {742}},\
  \bibinfo {pages} {200} (\bibinfo {year} {2015}{\natexlab{a}})},\ \Eprint
  {https://arxiv.org/abs/1409.3392} {arXiv:1409.3392 [nucl-ex]} \BibitemShut
  {NoStop}%
\bibitem [{\citenamefont {Khachatryan}\ \emph
  {et~al.}(2015{\natexlab{b}})\citenamefont {Khachatryan} \emph
  {et~al.}}]{Khachatryan:2015waa}%
  \BibitemOpen
  \bibfield  {author} {\bibinfo {author} {\bibfnamefont {V.}~\bibnamefont
  {Khachatryan}} \emph {et~al.} (\bibinfo {collaboration} {CMS}),\ }\bibfield
  {title} {\bibinfo {title} {{Evidence for collective multi-particle
  correlations in pPb collisions}},\ }\href
  {https://doi.org/10.1103/PhysRevLett.115.012301} {\bibfield  {journal}
  {\bibinfo  {journal} {Phys. Rev. Lett.}\ }\textbf {\bibinfo {volume} {115}},\
  \bibinfo {pages} {012301} (\bibinfo {year} {2015}{\natexlab{b}})},\ \Eprint
  {https://arxiv.org/abs/1502.05382} {arXiv:1502.05382 [nucl-ex]} \BibitemShut
  {NoStop}%
\bibitem [{\citenamefont {Aaboud}\ \emph {et~al.}(2017)\citenamefont {Aaboud}
  \emph {et~al.}}]{Aaboud:2017acw}%
  \BibitemOpen
  \bibfield  {author} {\bibinfo {author} {\bibfnamefont {M.}~\bibnamefont
  {Aaboud}} \emph {et~al.} (\bibinfo {collaboration} {ATLAS}),\ }\bibfield
  {title} {\bibinfo {title} {{Measurement of multi-particle azimuthal
  correlations in $pp$, $p+$Pb and low-multiplicity Pb$+$Pb collisions with the
  ATLAS detector}},\ }\href {https://doi.org/10.1140/epjc/s10052-017-4988-1}
  {\bibfield  {journal} {\bibinfo  {journal} {Eur. Phys. J. C}\ }\textbf
  {\bibinfo {volume} {77}},\ \bibinfo {pages} {428} (\bibinfo {year} {2017})},\
  \Eprint {https://arxiv.org/abs/1705.04176} {arXiv:1705.04176 [hep-ex]}
  \BibitemShut {NoStop}%
\bibitem [{\citenamefont {Aaboud}\ \emph {et~al.}(2018)\citenamefont {Aaboud}
  \emph {et~al.}}]{Aaboud:2017blb}%
  \BibitemOpen
  \bibfield  {author} {\bibinfo {author} {\bibfnamefont {M.}~\bibnamefont
  {Aaboud}} \emph {et~al.} (\bibinfo {collaboration} {ATLAS}),\ }\bibfield
  {title} {\bibinfo {title} {{Measurement of long-range multiparticle azimuthal
  correlations with the subevent cumulant method in $pp$ and $p + Pb$
  collisions with the ATLAS detector at the CERN Large Hadron Collider}},\
  }\href {https://doi.org/10.1103/PhysRevC.97.024904} {\bibfield  {journal}
  {\bibinfo  {journal} {Phys. Rev. C}\ }\textbf {\bibinfo {volume} {97}},\
  \bibinfo {pages} {024904} (\bibinfo {year} {2018})},\ \Eprint
  {https://arxiv.org/abs/1708.03559} {arXiv:1708.03559 [hep-ex]} \BibitemShut
  {NoStop}%
\bibitem [{\citenamefont {Aidala}\ \emph {et~al.}(2019)\citenamefont {Aidala}
  \emph {et~al.}}]{PHENIX:2018lia}%
  \BibitemOpen
  \bibfield  {author} {\bibinfo {author} {\bibfnamefont {C.}~\bibnamefont
  {Aidala}} \emph {et~al.} (\bibinfo {collaboration} {PHENIX}),\ }\bibfield
  {title} {\bibinfo {title} {{Creation of quark-gluon plasma droplets with
  three distinct geometries}},\ }\href
  {https://doi.org/10.1038/s41567-018-0360-0} {\bibfield  {journal} {\bibinfo
  {journal} {Nature Phys.}\ }\textbf {\bibinfo {volume} {15}},\ \bibinfo
  {pages} {214} (\bibinfo {year} {2019})},\ \Eprint
  {https://arxiv.org/abs/1805.02973} {arXiv:1805.02973 [nucl-ex]} \BibitemShut
  {NoStop}%
\bibitem [{\citenamefont {Adamczyk}\ \emph {et~al.}(2015)\citenamefont
  {Adamczyk} \emph {et~al.}}]{Adamczyk:2015xjc}%
  \BibitemOpen
  \bibfield  {author} {\bibinfo {author} {\bibfnamefont {L.}~\bibnamefont
  {Adamczyk}} \emph {et~al.} (\bibinfo {collaboration} {STAR}),\ }\bibfield
  {title} {\bibinfo {title} {{Long-range pseudorapidity dihadron correlations
  in $d$+Au collisions at $\sqrt{s_{\rm NN}}=200$ GeV}},\ }\href
  {https://doi.org/10.1016/j.physletb.2015.05.075} {\bibfield  {journal}
  {\bibinfo  {journal} {Phys. Lett. B}\ }\textbf {\bibinfo {volume} {747}},\
  \bibinfo {pages} {265} (\bibinfo {year} {2015})},\ \Eprint
  {https://arxiv.org/abs/1502.07652} {arXiv:1502.07652 [nucl-ex]} \BibitemShut
  {NoStop}%
\bibitem [{\citenamefont {Adare}\ \emph {et~al.}(2015)\citenamefont {Adare}
  \emph {et~al.}}]{Adare:2015ctn}%
  \BibitemOpen
  \bibfield  {author} {\bibinfo {author} {\bibfnamefont {A.}~\bibnamefont
  {Adare}} \emph {et~al.} (\bibinfo {collaboration} {PHENIX}),\ }\bibfield
  {title} {\bibinfo {title} {{Measurements of elliptic and triangular flow in
  high-multiplicity $^{3}$He$+$Au collisions at $\sqrt{s_{\rm NN}}$ = 200
  GeV}},\ }\href {https://doi.org/10.1103/PhysRevLett.115.142301} {\bibfield
  {journal} {\bibinfo  {journal} {Phys. Rev. Lett.}\ }\textbf {\bibinfo
  {volume} {115}},\ \bibinfo {pages} {142301} (\bibinfo {year} {2015})},\
  \Eprint {https://arxiv.org/abs/1507.06273} {arXiv:1507.06273 [nucl-ex]}
  \BibitemShut {NoStop}%
\bibitem [{\citenamefont {Aidala}\ \emph {et~al.}(2018)\citenamefont {Aidala}
  \emph {et~al.}}]{Aidala:2017ajz}%
  \BibitemOpen
  \bibfield  {author} {\bibinfo {author} {\bibfnamefont {C.}~\bibnamefont
  {Aidala}} \emph {et~al.} (\bibinfo {collaboration} {PHENIX}),\ }\bibfield
  {title} {\bibinfo {title} {{Measurements of Multiparticle Correlations in
  $d+\mathrm{Au}$ Collisions at 200, 62.4, 39, and 19.6 GeV and $p+\mathrm{Au}$
  Collisions at 200 GeV and Implications for Collective Behavior}},\ }\href
  {https://doi.org/10.1103/PhysRevLett.120.062302} {\bibfield  {journal}
  {\bibinfo  {journal} {Phys. Rev. Lett.}\ }\textbf {\bibinfo {volume} {120}},\
  \bibinfo {pages} {062302} (\bibinfo {year} {2018})},\ \Eprint
  {https://arxiv.org/abs/1707.06108} {arXiv:1707.06108 [nucl-ex]} \BibitemShut
  {NoStop}%
\bibitem [{\citenamefont {Dusling}\ \emph {et~al.}(2016)\citenamefont
  {Dusling}, \citenamefont {Li},\ and\ \citenamefont
  {Schenke}}]{Dusling:2015gta}%
  \BibitemOpen
  \bibfield  {author} {\bibinfo {author} {\bibfnamefont {K.}~\bibnamefont
  {Dusling}}, \bibinfo {author} {\bibfnamefont {W.}~\bibnamefont {Li}},\ and\
  \bibinfo {author} {\bibfnamefont {B.}~\bibnamefont {Schenke}},\ }\bibfield
  {title} {\bibinfo {title} {{Novel collective phenomena in high-energy
  proton-proton and proton-nucleus collisions}},\ }\href
  {https://doi.org/10.1142/S0218301316300022} {\bibfield  {journal} {\bibinfo
  {journal} {Int. J. Mod. Phys. E}\ }\textbf {\bibinfo {volume} {25}},\
  \bibinfo {pages} {1630002} (\bibinfo {year} {2016})},\ \Eprint
  {https://arxiv.org/abs/1509.07939} {arXiv:1509.07939 [nucl-ex]} \BibitemShut
  {NoStop}%
\bibitem [{\citenamefont {Nagle}\ and\ \citenamefont
  {Zajc}(2018)}]{Nagle:2018nvi}%
  \BibitemOpen
  \bibfield  {author} {\bibinfo {author} {\bibfnamefont {J.~L.}\ \bibnamefont
  {Nagle}}\ and\ \bibinfo {author} {\bibfnamefont {W.~A.}\ \bibnamefont
  {Zajc}},\ }\bibfield  {title} {\bibinfo {title} {{Small System Collectivity
  in Relativistic Hadronic and Nuclear Collisions}},\ }\href
  {https://doi.org/10.1146/annurev-nucl-101916-123209} {\bibfield  {journal}
  {\bibinfo  {journal} {Ann. Rev. Nucl. Part. Sci.}\ }\textbf {\bibinfo
  {volume} {68}},\ \bibinfo {pages} {211} (\bibinfo {year} {2018})},\ \Eprint
  {https://arxiv.org/abs/1801.03477} {arXiv:1801.03477 [nucl-ex]} \BibitemShut
  {NoStop}%
\bibitem [{\citenamefont {Schenke}\ \emph {et~al.}(2020)\citenamefont
  {Schenke}, \citenamefont {Shen},\ and\ \citenamefont
  {Tribedy}}]{Schenke:2020mbo}%
  \BibitemOpen
  \bibfield  {author} {\bibinfo {author} {\bibfnamefont {B.}~\bibnamefont
  {Schenke}}, \bibinfo {author} {\bibfnamefont {C.}~\bibnamefont {Shen}},\ and\
  \bibinfo {author} {\bibfnamefont {P.}~\bibnamefont {Tribedy}},\ }\bibfield
  {title} {\bibinfo {title} {{Running the gamut of high energy nuclear
  collisions}},\ }\href {https://doi.org/10.1103/PhysRevC.102.044905}
  {\bibfield  {journal} {\bibinfo  {journal} {Phys. Rev. C}\ }\textbf {\bibinfo
  {volume} {102}},\ \bibinfo {pages} {044905} (\bibinfo {year} {2020})},\
  \Eprint {https://arxiv.org/abs/2005.14682} {arXiv:2005.14682 [nucl-th]}
  \BibitemShut {NoStop}%
\bibitem [{\citenamefont {Becattini}(1996)}]{Becattini:1995if}%
  \BibitemOpen
  \bibfield  {author} {\bibinfo {author} {\bibfnamefont {F.}~\bibnamefont
  {Becattini}},\ }\bibfield  {title} {\bibinfo {title} {{A Thermodynamical
  approach to hadron production in e+e- collisions}},\ }\href
  {https://doi.org/10.1007/BF02907431} {\bibfield  {journal} {\bibinfo
  {journal} {Z. Phys. C}\ }\textbf {\bibinfo {volume} {69}},\ \bibinfo {pages}
  {485} (\bibinfo {year} {1996})}\BibitemShut {NoStop}%
\bibitem [{\citenamefont {Becattini}\ \emph {et~al.}(2008)\citenamefont
  {Becattini}, \citenamefont {Castorina}, \citenamefont {Manninen},\ and\
  \citenamefont {Satz}}]{Becattini:2008tx}%
  \BibitemOpen
  \bibfield  {author} {\bibinfo {author} {\bibfnamefont {F.}~\bibnamefont
  {Becattini}}, \bibinfo {author} {\bibfnamefont {P.}~\bibnamefont
  {Castorina}}, \bibinfo {author} {\bibfnamefont {J.}~\bibnamefont
  {Manninen}},\ and\ \bibinfo {author} {\bibfnamefont {H.}~\bibnamefont
  {Satz}},\ }\bibfield  {title} {\bibinfo {title} {{The Thermal Production of
  Strange and Non-Strange Hadrons in e+ e- Collisions}},\ }\href
  {https://doi.org/10.1140/epjc/s10052-008-0671-x} {\bibfield  {journal}
  {\bibinfo  {journal} {Eur. Phys. J. C}\ }\textbf {\bibinfo {volume} {56}},\
  \bibinfo {pages} {493} (\bibinfo {year} {2008})},\ \Eprint
  {https://arxiv.org/abs/0805.0964} {arXiv:0805.0964 [hep-ph]} \BibitemShut
  {NoStop}%
\bibitem [{\citenamefont {Castorina}\ \emph {et~al.}(2007)\citenamefont
  {Castorina}, \citenamefont {Kharzeev},\ and\ \citenamefont
  {Satz}}]{Castorina:2007eb}%
  \BibitemOpen
  \bibfield  {author} {\bibinfo {author} {\bibfnamefont {P.}~\bibnamefont
  {Castorina}}, \bibinfo {author} {\bibfnamefont {D.}~\bibnamefont
  {Kharzeev}},\ and\ \bibinfo {author} {\bibfnamefont {H.}~\bibnamefont
  {Satz}},\ }\bibfield  {title} {\bibinfo {title} {{Thermal Hadronization and
  Hawking-Unruh Radiation in QCD}},\ }\href
  {https://doi.org/10.1140/epjc/s10052-007-0368-6} {\bibfield  {journal}
  {\bibinfo  {journal} {Eur. Phys. J. C}\ }\textbf {\bibinfo {volume} {52}},\
  \bibinfo {pages} {187} (\bibinfo {year} {2007})},\ \Eprint
  {https://arxiv.org/abs/0704.1426} {arXiv:0704.1426 [hep-ph]} \BibitemShut
  {NoStop}%
\bibitem [{\citenamefont {Braun-Munzinger}\ \emph {et~al.}(1995)\citenamefont
  {Braun-Munzinger}, \citenamefont {Stachel}, \citenamefont {Wessels},\ and\
  \citenamefont {Xu}}]{BraunMunzinger:1994xr}%
  \BibitemOpen
  \bibfield  {author} {\bibinfo {author} {\bibfnamefont {P.}~\bibnamefont
  {Braun-Munzinger}}, \bibinfo {author} {\bibfnamefont {J.}~\bibnamefont
  {Stachel}}, \bibinfo {author} {\bibfnamefont {J.~P.}\ \bibnamefont
  {Wessels}},\ and\ \bibinfo {author} {\bibfnamefont {N.}~\bibnamefont {Xu}},\
  }\bibfield  {title} {\bibinfo {title} {{Thermal equilibration and expansion
  in nucleus-nucleus collisions at the AGS}},\ }\href
  {https://doi.org/10.1016/0370-2693(94)01534-J} {\bibfield  {journal}
  {\bibinfo  {journal} {Phys. Lett. B}\ }\textbf {\bibinfo {volume} {344}},\
  \bibinfo {pages} {43} (\bibinfo {year} {1995})},\ \Eprint
  {https://arxiv.org/abs/nucl-th/9410026} {arXiv:nucl-th/9410026 [nucl-th]}
  \BibitemShut {NoStop}%
\bibitem [{\citenamefont {Hawking}(1975)}]{Hawking:1974sw}%
  \BibitemOpen
  \bibfield  {author} {\bibinfo {author} {\bibfnamefont {S.}~\bibnamefont
  {Hawking}},\ }\bibfield  {title} {\bibinfo {title} {{Particle Creation by
  Black Holes}},\ }\href {https://doi.org/10.1007/BF02345020} {\bibfield
  {journal} {\bibinfo  {journal} {Commun.\ Math.\ Phys.}\ }\textbf {\bibinfo
  {volume} {43}},\ \bibinfo {pages} {199} (\bibinfo {year} {1975})},\ \bibinfo
  {note} {[Erratum: Commun.Math.Phys. 46, 206 (1976)]}\BibitemShut {NoStop}%
\bibitem [{\citenamefont {Unruh}(1976)}]{Unruh:1976db}%
  \BibitemOpen
  \bibfield  {author} {\bibinfo {author} {\bibfnamefont {W.}~\bibnamefont
  {Unruh}},\ }\bibfield  {title} {\bibinfo {title} {{Notes on black hole
  evaporation}},\ }\href {https://doi.org/10.1103/PhysRevD.14.870} {\bibfield
  {journal} {\bibinfo  {journal} {Phys.\ Rev.\ D}\ }\textbf {\bibinfo {volume}
  {14}},\ \bibinfo {pages} {870} (\bibinfo {year} {1976})}\BibitemShut
  {NoStop}%
\bibitem [{\citenamefont {Kharzeev}\ and\ \citenamefont
  {Levin}(2017)}]{Kharzeev:2017qzs}%
  \BibitemOpen
  \bibfield  {author} {\bibinfo {author} {\bibfnamefont {D.~E.}\ \bibnamefont
  {Kharzeev}}\ and\ \bibinfo {author} {\bibfnamefont {E.~M.}\ \bibnamefont
  {Levin}},\ }\bibfield  {title} {\bibinfo {title} {{Deep inelastic scattering
  as a probe of entanglement}},\ }\href
  {https://doi.org/10.1103/PhysRevD.95.114008} {\bibfield  {journal} {\bibinfo
  {journal} {Phys. Rev. D}\ }\textbf {\bibinfo {volume} {95}},\ \bibinfo
  {pages} {114008} (\bibinfo {year} {2017})},\ \Eprint
  {https://arxiv.org/abs/1702.03489} {arXiv:1702.03489 [hep-ph]} \BibitemShut
  {NoStop}%
\bibitem [{\citenamefont {Berges}\ \emph {et~al.}(2018)\citenamefont {Berges},
  \citenamefont {Floerchinger},\ and\ \citenamefont
  {Venugopalan}}]{Berges:2017zws}%
  \BibitemOpen
  \bibfield  {author} {\bibinfo {author} {\bibfnamefont {J.}~\bibnamefont
  {Berges}}, \bibinfo {author} {\bibfnamefont {S.}~\bibnamefont
  {Floerchinger}},\ and\ \bibinfo {author} {\bibfnamefont {R.}~\bibnamefont
  {Venugopalan}},\ }\bibfield  {title} {\bibinfo {title} {{Thermal excitation
  spectrum from entanglement in an expanding quantum string}},\ }\href
  {https://doi.org/10.1016/j.physletb.2018.01.068} {\bibfield  {journal}
  {\bibinfo  {journal} {Phys. Lett. B}\ }\textbf {\bibinfo {volume} {778}},\
  \bibinfo {pages} {442} (\bibinfo {year} {2018})},\ \Eprint
  {https://arxiv.org/abs/1707.05338} {arXiv:1707.05338 [hep-ph]} \BibitemShut
  {NoStop}%
\bibitem [{\citenamefont {Kaufman}\ \emph {et~al.}(2016)\citenamefont
  {Kaufman}, \citenamefont {Tai}, \citenamefont {Lukin}, \citenamefont
  {Rispoli}, \citenamefont {Schittko}, \citenamefont {Preiss},\ and\
  \citenamefont {Greiner}}]{Kaufman_2016}%
  \BibitemOpen
  \bibfield  {author} {\bibinfo {author} {\bibfnamefont {A.~M.}\ \bibnamefont
  {Kaufman}}, \bibinfo {author} {\bibfnamefont {M.~E.}\ \bibnamefont {Tai}},
  \bibinfo {author} {\bibfnamefont {A.}~\bibnamefont {Lukin}}, \bibinfo
  {author} {\bibfnamefont {M.}~\bibnamefont {Rispoli}}, \bibinfo {author}
  {\bibfnamefont {R.}~\bibnamefont {Schittko}}, \bibinfo {author}
  {\bibfnamefont {P.~M.}\ \bibnamefont {Preiss}},\ and\ \bibinfo {author}
  {\bibfnamefont {M.}~\bibnamefont {Greiner}},\ }\bibfield  {title} {\bibinfo
  {title} {Quantum thermalization through entanglement in an isolated many-body
  system},\ }\href {https://doi.org/10.1126/science.aaf6725} {\bibfield
  {journal} {\bibinfo  {journal} {Science}\ }\textbf {\bibinfo {volume}
  {353}},\ \bibinfo {pages} {794} (\bibinfo {year} {2016})}\BibitemShut
  {NoStop}%
\bibitem [{\citenamefont {Baker}\ and\ \citenamefont
  {Kharzeev}(2018)}]{Baker:2017wtt}%
  \BibitemOpen
  \bibfield  {author} {\bibinfo {author} {\bibfnamefont {O.~K.}\ \bibnamefont
  {Baker}}\ and\ \bibinfo {author} {\bibfnamefont {D.~E.}\ \bibnamefont
  {Kharzeev}},\ }\bibfield  {title} {\bibinfo {title} {{Thermal radiation and
  entanglement in proton-proton collisions at energies available at the CERN
  Large Hadron Collider}},\ }\href {https://doi.org/10.1103/PhysRevD.98.054007}
  {\bibfield  {journal} {\bibinfo  {journal} {Phys. Rev. D}\ }\textbf {\bibinfo
  {volume} {98}},\ \bibinfo {pages} {054007} (\bibinfo {year} {2018})},\
  \Eprint {https://arxiv.org/abs/1712.04558} {arXiv:1712.04558 [hep-ph]}
  \BibitemShut {NoStop}%
\bibitem [{\citenamefont {Tu}\ \emph {et~al.}(2020)\citenamefont {Tu},
  \citenamefont {Kharzeev},\ and\ \citenamefont {Ullrich}}]{Tu:2019ouv}%
  \BibitemOpen
  \bibfield  {author} {\bibinfo {author} {\bibfnamefont {Z.}~\bibnamefont
  {Tu}}, \bibinfo {author} {\bibfnamefont {D.~E.}\ \bibnamefont {Kharzeev}},\
  and\ \bibinfo {author} {\bibfnamefont {T.}~\bibnamefont {Ullrich}},\
  }\bibfield  {title} {\bibinfo {title} {{The EPR paradox and quantum
  entanglement at sub-nucleonic scales}},\ }\href
  {https://doi.org/10.1103/PhysRevLett.124.062001} {\bibfield  {journal}
  {\bibinfo  {journal} {Phys. Rev. Lett.}\ }\textbf {\bibinfo {volume} {124}},\
  \bibinfo {pages} {062001} (\bibinfo {year} {2020})},\ \Eprint
  {https://arxiv.org/abs/1904.11974} {arXiv:1904.11974 [hep-ph]} \BibitemShut
  {NoStop}%
\bibitem [{\citenamefont {Badea}\ \emph {et~al.}(2019)\citenamefont {Badea},
  \citenamefont {Baty}, \citenamefont {Chang}, \citenamefont {Innocenti},
  \citenamefont {Maggi}, \citenamefont {Mcginn}, \citenamefont {Peters},
  \citenamefont {Sheng}, \citenamefont {Thaler},\ and\ \citenamefont
  {Lee}}]{Badea:2019vey}%
  \BibitemOpen
  \bibfield  {author} {\bibinfo {author} {\bibfnamefont {A.}~\bibnamefont
  {Badea}}, \bibinfo {author} {\bibfnamefont {A.}~\bibnamefont {Baty}},
  \bibinfo {author} {\bibfnamefont {P.}~\bibnamefont {Chang}}, \bibinfo
  {author} {\bibfnamefont {G.~M.}\ \bibnamefont {Innocenti}}, \bibinfo {author}
  {\bibfnamefont {M.}~\bibnamefont {Maggi}}, \bibinfo {author} {\bibfnamefont
  {C.}~\bibnamefont {Mcginn}}, \bibinfo {author} {\bibfnamefont
  {M.}~\bibnamefont {Peters}}, \bibinfo {author} {\bibfnamefont {T.-A.}\
  \bibnamefont {Sheng}}, \bibinfo {author} {\bibfnamefont {J.}~\bibnamefont
  {Thaler}},\ and\ \bibinfo {author} {\bibfnamefont {Y.-J.}\ \bibnamefont
  {Lee}},\ }\bibfield  {title} {\bibinfo {title} {{Measurements of two-particle
  correlations in $e^+e^-$ collisions at 91 GeV with ALEPH archived data}},\
  }\href {https://doi.org/10.1103/PhysRevLett.123.212002} {\bibfield  {journal}
  {\bibinfo  {journal} {Phys. Rev. Lett.}\ }\textbf {\bibinfo {volume} {123}},\
  \bibinfo {pages} {212002} (\bibinfo {year} {2019})},\ \Eprint
  {https://arxiv.org/abs/1906.00489} {arXiv:1906.00489 [hep-ex]} \BibitemShut
  {NoStop}%
\bibitem [{\citenamefont {Abt}\ \emph {et~al.}(2019)\citenamefont {Abt} \emph
  {et~al.}}]{ZEUS:2019jya}%
  \BibitemOpen
  \bibfield  {author} {\bibinfo {author} {\bibfnamefont {I.}~\bibnamefont
  {Abt}} \emph {et~al.} (\bibinfo {collaboration} {ZEUS}),\ }\bibfield  {title}
  {\bibinfo {title} {{Two-particle azimuthal correlations as a probe of
  collective behaviour in deep inelastic $\textit{ep}$ scattering at HERA}},\
  }\Eprint {https://arxiv.org/abs/1912.07431} {arXiv:1912.07431 [hep-ex]}
  (\bibinfo {year} {2019})\BibitemShut {NoStop}%
\bibitem [{Note1()}]{Note1}%
  \BibitemOpen
  \bibinfo {note} {``QGP-like'' refers to the state where qualitative
  signatures of partonic collectivity are present but the system does not
  necessarily reach the hydrodynamic limit.}\BibitemShut {Stop}%
\bibitem [{\citenamefont {Mueller}(1983)}]{Mueller:1982cq}%
  \BibitemOpen
  \bibfield  {author} {\bibinfo {author} {\bibfnamefont {A.~H.}\ \bibnamefont
  {Mueller}},\ }\bibfield  {title} {\bibinfo {title} {{Multiplicity and Hadron
  Distributions in QCD Jets: Nonleading Terms}},\ }\href
  {https://doi.org/10.1016/0550-3213(83)90176-1} {\bibfield  {journal}
  {\bibinfo  {journal} {Nucl. Phys. B}\ }\textbf {\bibinfo {volume} {213}},\
  \bibinfo {pages} {85} (\bibinfo {year} {1983})}\BibitemShut {NoStop}%
\bibitem [{\citenamefont {Cacciari}\ \emph {et~al.}(2008)\citenamefont
  {Cacciari}, \citenamefont {Salam},\ and\ \citenamefont
  {Soyez}}]{Cacciari:2008gp}%
  \BibitemOpen
  \bibfield  {author} {\bibinfo {author} {\bibfnamefont {M.}~\bibnamefont
  {Cacciari}}, \bibinfo {author} {\bibfnamefont {G.~P.}\ \bibnamefont
  {Salam}},\ and\ \bibinfo {author} {\bibfnamefont {G.}~\bibnamefont {Soyez}},\
  }\bibfield  {title} {\bibinfo {title} {{The anti-$k_t$ jet clustering
  algorithm}},\ }\href {https://doi.org/10.1088/1126-6708/2008/04/063}
  {\bibfield  {journal} {\bibinfo  {journal} {JHEP}\ }\textbf {\bibinfo
  {volume} {04}},\ \bibinfo {pages} {063}},\ \Eprint
  {https://arxiv.org/abs/0802.1189} {arXiv:0802.1189 [hep-ph]} \BibitemShut
  {NoStop}%
\bibitem [{\citenamefont {Sjostrand}\ \emph {et~al.}(2008)\citenamefont
  {Sjostrand}, \citenamefont {Mrenna},\ and\ \citenamefont
  {Skands}}]{Sjostrand:2007gs}%
  \BibitemOpen
  \bibfield  {author} {\bibinfo {author} {\bibfnamefont {T.}~\bibnamefont
  {Sjostrand}}, \bibinfo {author} {\bibfnamefont {S.}~\bibnamefont {Mrenna}},\
  and\ \bibinfo {author} {\bibfnamefont {P.~Z.}\ \bibnamefont {Skands}},\
  }\bibfield  {title} {\bibinfo {title} {{A Brief Introduction to PYTHIA
  8.1}},\ }\href {https://doi.org/10.1016/j.cpc.2008.01.036} {\bibfield
  {journal} {\bibinfo  {journal} {Comput. Phys. Commun.}\ }\textbf {\bibinfo
  {volume} {178}},\ \bibinfo {pages} {852} (\bibinfo {year} {2008})},\ \Eprint
  {https://arxiv.org/abs/0710.3820} {arXiv:0710.3820 [hep-ph]} \BibitemShut
  {NoStop}%
\bibitem [{\citenamefont {Aad}\ \emph {et~al.}(2016{\natexlab{b}})\citenamefont
  {Aad} \emph {et~al.}}]{Aad:2016oit}%
  \BibitemOpen
  \bibfield  {author} {\bibinfo {author} {\bibfnamefont {G.}~\bibnamefont
  {Aad}} \emph {et~al.} (\bibinfo {collaboration} {ATLAS}),\ }\bibfield
  {title} {\bibinfo {title} {{Measurement of the charged-particle multiplicity
  inside jets from $\sqrt{s}=8$ TeV $pp$ collisions with the ATLAS detector}},\
  }\href {https://doi.org/10.1140/epjc/s10052-016-4126-5} {\bibfield  {journal}
  {\bibinfo  {journal} {Eur. Phys. J. C}\ }\textbf {\bibinfo {volume} {76}},\
  \bibinfo {pages} {322} (\bibinfo {year} {2016}{\natexlab{b}})},\ \Eprint
  {https://arxiv.org/abs/1602.00988} {arXiv:1602.00988 [hep-ex]} \BibitemShut
  {NoStop}%
\bibitem [{\citenamefont {Koch}\ \emph {et~al.}(1986)\citenamefont {Koch},
  \citenamefont {Muller},\ and\ \citenamefont {Rafelski}}]{Koch:1986ud}%
  \BibitemOpen
  \bibfield  {author} {\bibinfo {author} {\bibfnamefont {P.}~\bibnamefont
  {Koch}}, \bibinfo {author} {\bibfnamefont {B.}~\bibnamefont {Muller}},\ and\
  \bibinfo {author} {\bibfnamefont {J.}~\bibnamefont {Rafelski}},\ }\bibfield
  {title} {\bibinfo {title} {{Strangeness in Relativistic Heavy Ion
  Collisions}},\ }\href {https://doi.org/10.1016/0370-1573(86)90096-7}
  {\bibfield  {journal} {\bibinfo  {journal} {Phys. Rept.}\ }\textbf {\bibinfo
  {volume} {142}},\ \bibinfo {pages} {167} (\bibinfo {year}
  {1986})}\BibitemShut {NoStop}%
\bibitem [{\citenamefont {Adam}\ \emph {et~al.}(2017)\citenamefont {Adam} \emph
  {et~al.}}]{ALICE:2017jyt}%
  \BibitemOpen
  \bibfield  {author} {\bibinfo {author} {\bibfnamefont {J.}~\bibnamefont
  {Adam}} \emph {et~al.} (\bibinfo {collaboration} {ALICE}),\ }\bibfield
  {title} {\bibinfo {title} {{Enhanced production of multi-strange hadrons in
  high-multiplicity proton-proton collisions}},\ }\href
  {https://doi.org/10.1038/nphys4111} {\bibfield  {journal} {\bibinfo
  {journal} {Nature Phys.}\ }\textbf {\bibinfo {volume} {13}},\ \bibinfo
  {pages} {535} (\bibinfo {year} {2017})},\ \Eprint
  {https://arxiv.org/abs/1606.07424} {arXiv:1606.07424 [nucl-ex]} \BibitemShut
  {NoStop}%
\bibitem [{\citenamefont {Gatoff}\ and\ \citenamefont
  {Wong}(1992)}]{Gatoff:1992cv}%
  \BibitemOpen
  \bibfield  {author} {\bibinfo {author} {\bibfnamefont {G.}~\bibnamefont
  {Gatoff}}\ and\ \bibinfo {author} {\bibfnamefont {C.~Y.}\ \bibnamefont
  {Wong}},\ }\bibfield  {title} {\bibinfo {title} {{Origin of the soft p(T)
  spectra}},\ }\href {https://doi.org/10.1103/PhysRevD.46.997} {\bibfield
  {journal} {\bibinfo  {journal} {Phys. Rev. D}\ }\textbf {\bibinfo {volume}
  {46}},\ \bibinfo {pages} {997} (\bibinfo {year} {1992})}\BibitemShut
  {NoStop}%
\bibitem [{\citenamefont {Hagedorn}(1965)}]{Hagedorn:1965st}%
  \BibitemOpen
  \bibfield  {author} {\bibinfo {author} {\bibfnamefont {R.}~\bibnamefont
  {Hagedorn}},\ }\bibfield  {title} {\bibinfo {title} {{Statistical
  thermodynamics of strong interactions at high-energies}},\ }\href@noop {}
  {\bibfield  {journal} {\bibinfo  {journal} {Nuovo Cim. Suppl.}\ }\textbf
  {\bibinfo {volume} {3}},\ \bibinfo {pages} {147} (\bibinfo {year}
  {1965})}\BibitemShut {NoStop}%
\bibitem [{\citenamefont {Khachatryan}\ \emph
  {et~al.}(2017{\natexlab{b}})\citenamefont {Khachatryan} \emph
  {et~al.}}]{Khachatryan:2016yru}%
  \BibitemOpen
  \bibfield  {author} {\bibinfo {author} {\bibfnamefont {V.}~\bibnamefont
  {Khachatryan}} \emph {et~al.} (\bibinfo {collaboration} {CMS}),\ }\bibfield
  {title} {\bibinfo {title} {{Multiplicity and rapidity dependence of strange
  hadron production in pp, pPb, and PbPb collisions at the LHC}},\ }\href
  {https://doi.org/10.1016/j.physletb.2017.01.075} {\bibfield  {journal}
  {\bibinfo  {journal} {Phys. Lett. B}\ }\textbf {\bibinfo {volume} {768}},\
  \bibinfo {pages} {103} (\bibinfo {year} {2017}{\natexlab{b}})},\ \Eprint
  {https://arxiv.org/abs/1605.06699} {arXiv:1605.06699 [nucl-ex]} \BibitemShut
  {NoStop}%
\bibitem [{\citenamefont {Acharya}\ \emph {et~al.}(2020)\citenamefont {Acharya}
  \emph {et~al.}}]{Acharya:2020zji}%
  \BibitemOpen
  \bibfield  {author} {\bibinfo {author} {\bibfnamefont {S.}~\bibnamefont
  {Acharya}} \emph {et~al.} (\bibinfo {collaboration} {ALICE}),\ }\bibfield
  {title} {\bibinfo {title} {{Multiplicity dependence of $\pi $, K, and p
  production in pp collisions at $\sqrt{s} = 13$ TeV}},\ }\href
  {https://doi.org/10.1140/epjc/s10052-020-8125-1} {\bibfield  {journal}
  {\bibinfo  {journal} {Eur. Phys. J. C}\ }\textbf {\bibinfo {volume} {80}},\
  \bibinfo {pages} {693} (\bibinfo {year} {2020})},\ \Eprint
  {https://arxiv.org/abs/2003.02394} {arXiv:2003.02394 [nucl-ex]} \BibitemShut
  {NoStop}%
\bibitem [{\citenamefont {Abelev}\ \emph {et~al.}(2014)\citenamefont {Abelev}
  \emph {et~al.}}]{Abelev:2013haa}%
  \BibitemOpen
  \bibfield  {author} {\bibinfo {author} {\bibfnamefont {B.~B.}\ \bibnamefont
  {Abelev}} \emph {et~al.} (\bibinfo {collaboration} {ALICE}),\ }\bibfield
  {title} {\bibinfo {title} {{Multiplicity Dependence of Pion, Kaon, Proton and
  Lambda Production in p-Pb Collisions at $\sqrt{s_{NN}}$ = 5.02 TeV}},\ }\href
  {https://doi.org/10.1016/j.physletb.2013.11.020} {\bibfield  {journal}
  {\bibinfo  {journal} {Phys. Lett. B}\ }\textbf {\bibinfo {volume} {728}},\
  \bibinfo {pages} {25} (\bibinfo {year} {2014})},\ \Eprint
  {https://arxiv.org/abs/1307.6796} {arXiv:1307.6796 [nucl-ex]} \BibitemShut
  {NoStop}%
\bibitem [{\citenamefont {Hanbury~Brown}\ and\ \citenamefont
  {Twiss}(1956)}]{HanburyBrown:1956bqd}%
  \BibitemOpen
  \bibfield  {author} {\bibinfo {author} {\bibfnamefont {R.}~\bibnamefont
  {Hanbury~Brown}}\ and\ \bibinfo {author} {\bibfnamefont {R.~Q.}\ \bibnamefont
  {Twiss}},\ }\bibfield  {title} {\bibinfo {title} {{A Test of a new type of
  stellar interferometer on Sirius}},\ }\href
  {https://doi.org/10.1038/1781046a0} {\bibfield  {journal} {\bibinfo
  {journal} {Nature}\ }\textbf {\bibinfo {volume} {178}},\ \bibinfo {pages}
  {1046} (\bibinfo {year} {1956})}\BibitemShut {NoStop}%
\bibitem [{\citenamefont {Zajc}\ \emph {et~al.}(1984)\citenamefont {Zajc} \emph
  {et~al.}}]{Zajc:1984vb}%
  \BibitemOpen
  \bibfield  {author} {\bibinfo {author} {\bibfnamefont {W.~A.}\ \bibnamefont
  {Zajc}} \emph {et~al.},\ }\bibfield  {title} {\bibinfo {title} {{Two pion
  correlations in heavy ion collisions}},\ }\href
  {https://doi.org/10.1103/PhysRevC.29.2173} {\bibfield  {journal} {\bibinfo
  {journal} {Phys. Rev. C}\ }\textbf {\bibinfo {volume} {29}},\ \bibinfo
  {pages} {2173} (\bibinfo {year} {1984})}\BibitemShut {NoStop}%
\bibitem [{\citenamefont {Heinz}\ and\ \citenamefont
  {Jacak}(1999)}]{Heinz:1999rw}%
  \BibitemOpen
  \bibfield  {author} {\bibinfo {author} {\bibfnamefont {U.~W.}\ \bibnamefont
  {Heinz}}\ and\ \bibinfo {author} {\bibfnamefont {B.~V.}\ \bibnamefont
  {Jacak}},\ }\bibfield  {title} {\bibinfo {title} {{Two particle correlations
  in relativistic heavy ion collisions}},\ }\href
  {https://doi.org/10.1146/annurev.nucl.49.1.529} {\bibfield  {journal}
  {\bibinfo  {journal} {Ann. Rev. Nucl. Part. Sci.}\ }\textbf {\bibinfo
  {volume} {49}},\ \bibinfo {pages} {529} (\bibinfo {year} {1999})},\ \Eprint
  {https://arxiv.org/abs/nucl-th/9902020} {arXiv:nucl-th/9902020} \BibitemShut
  {NoStop}%
\bibitem [{\citenamefont {Lisa}\ \emph {et~al.}(2005)\citenamefont {Lisa},
  \citenamefont {Pratt}, \citenamefont {Soltz},\ and\ \citenamefont
  {Wiedemann}}]{Lisa:2005dd}%
  \BibitemOpen
  \bibfield  {author} {\bibinfo {author} {\bibfnamefont {M.~A.}\ \bibnamefont
  {Lisa}}, \bibinfo {author} {\bibfnamefont {S.}~\bibnamefont {Pratt}},
  \bibinfo {author} {\bibfnamefont {R.}~\bibnamefont {Soltz}},\ and\ \bibinfo
  {author} {\bibfnamefont {U.}~\bibnamefont {Wiedemann}},\ }\bibfield  {title}
  {\bibinfo {title} {{Femtoscopy in relativistic heavy ion collisions}},\
  }\href {https://doi.org/10.1146/annurev.nucl.55.090704.151533} {\bibfield
  {journal} {\bibinfo  {journal} {Ann. Rev. Nucl. Part. Sci.}\ }\textbf
  {\bibinfo {volume} {55}},\ \bibinfo {pages} {357} (\bibinfo {year} {2005})},\
  \Eprint {https://arxiv.org/abs/nucl-ex/0505014} {arXiv:nucl-ex/0505014}
  \BibitemShut {NoStop}%
\bibitem [{\citenamefont {Adare}\ \emph {et~al.}(2010)\citenamefont {Adare}
  \emph {et~al.}}]{Adare:2008ab}%
  \BibitemOpen
  \bibfield  {author} {\bibinfo {author} {\bibfnamefont {A.}~\bibnamefont
  {Adare}} \emph {et~al.} (\bibinfo {collaboration} {PHENIX}),\ }\bibfield
  {title} {\bibinfo {title} {{Enhanced production of direct photons in Au+Au
  collisions at $\sqrt{s_{NN}}=200$ GeV and implications for the initial
  temperature}},\ }\href {https://doi.org/10.1103/PhysRevLett.104.132301}
  {\bibfield  {journal} {\bibinfo  {journal} {Phys. Rev. Lett.}\ }\textbf
  {\bibinfo {volume} {104}},\ \bibinfo {pages} {132301} (\bibinfo {year}
  {2010})},\ \Eprint {https://arxiv.org/abs/0804.4168} {arXiv:0804.4168
  [nucl-ex]} \BibitemShut {NoStop}%
\bibitem [{\citenamefont {Adam}\ \emph {et~al.}(2016)\citenamefont {Adam} \emph
  {et~al.}}]{Adam:2015lda}%
  \BibitemOpen
  \bibfield  {author} {\bibinfo {author} {\bibfnamefont {J.}~\bibnamefont
  {Adam}} \emph {et~al.} (\bibinfo {collaboration} {ALICE}),\ }\bibfield
  {title} {\bibinfo {title} {{Direct photon production in Pb-Pb collisions at
  $\sqrt{s_{\rm{NN}}} =$ 2.76 TeV}},\ }\href
  {https://doi.org/10.1016/j.physletb.2016.01.020} {\bibfield  {journal}
  {\bibinfo  {journal} {Phys. Lett. B}\ }\textbf {\bibinfo {volume} {754}},\
  \bibinfo {pages} {235} (\bibinfo {year} {2016})},\ \Eprint
  {https://arxiv.org/abs/1509.07324} {arXiv:1509.07324 [nucl-ex]} \BibitemShut
  {NoStop}%
\bibitem [{\citenamefont {Adamczyk}\ \emph {et~al.}(2017)\citenamefont
  {Adamczyk} \emph {et~al.}}]{STAR:2016use}%
  \BibitemOpen
  \bibfield  {author} {\bibinfo {author} {\bibfnamefont {L.}~\bibnamefont
  {Adamczyk}} \emph {et~al.} (\bibinfo {collaboration} {STAR}),\ }\bibfield
  {title} {\bibinfo {title} {{Direct virtual photon production in Au+Au
  collisions at $\sqrt{s_{NN}}$ = 200 GeV}},\ }\href
  {https://doi.org/10.1016/j.physletb.2017.04.050} {\bibfield  {journal}
  {\bibinfo  {journal} {Phys. Lett. B}\ }\textbf {\bibinfo {volume} {770}},\
  \bibinfo {pages} {451} (\bibinfo {year} {2017})},\ \Eprint
  {https://arxiv.org/abs/1607.01447} {arXiv:1607.01447 [nucl-ex]} \BibitemShut
  {NoStop}%
\bibitem [{\citenamefont {Bertolini}\ \emph {et~al.}(2014)\citenamefont
  {Bertolini}, \citenamefont {Chan},\ and\ \citenamefont
  {Thaler}}]{Bertolini:2013iqa}%
  \BibitemOpen
  \bibfield  {author} {\bibinfo {author} {\bibfnamefont {D.}~\bibnamefont
  {Bertolini}}, \bibinfo {author} {\bibfnamefont {T.}~\bibnamefont {Chan}},\
  and\ \bibinfo {author} {\bibfnamefont {J.}~\bibnamefont {Thaler}},\
  }\bibfield  {title} {\bibinfo {title} {{Jet Observables Without Jet
  Algorithms}},\ }\href {https://doi.org/10.1007/JHEP04(2014)013} {\bibfield
  {journal} {\bibinfo  {journal} {JHEP}\ }\textbf {\bibinfo {volume} {04}},\
  \bibinfo {pages} {013}},\ \Eprint {https://arxiv.org/abs/1310.7584}
  {arXiv:1310.7584 [hep-ph]} \BibitemShut {NoStop}%
\bibitem [{\citenamefont {Larkoski}\ \emph {et~al.}(2014)\citenamefont
  {Larkoski}, \citenamefont {Neill},\ and\ \citenamefont
  {Thaler}}]{Larkoski:2014uqa}%
  \BibitemOpen
  \bibfield  {author} {\bibinfo {author} {\bibfnamefont {A.~J.}\ \bibnamefont
  {Larkoski}}, \bibinfo {author} {\bibfnamefont {D.}~\bibnamefont {Neill}},\
  and\ \bibinfo {author} {\bibfnamefont {J.}~\bibnamefont {Thaler}},\
  }\bibfield  {title} {\bibinfo {title} {{Jet Shapes with the Broadening
  Axis}},\ }\href {https://doi.org/10.1007/JHEP04(2014)017} {\bibfield
  {journal} {\bibinfo  {journal} {JHEP}\ }\textbf {\bibinfo {volume} {04}},\
  \bibinfo {pages} {017}},\ \Eprint {https://arxiv.org/abs/1401.2158}
  {arXiv:1401.2158 [hep-ph]} \BibitemShut {NoStop}%
\end{thebibliography}%
\end{document}